\documentstyle[preprint,aps,eqsecnum, epsfig]{revtex}
\begin{document}
\tighten
\preprint{PITT-97-1; CMU-HEP-97-03; DOR-ER/40682-128;LPTHE-97-}
\draft 
\title{PHOTOPRODUCTION  ENHANCEMENT FROM NON EQUILIBRIUM DISORIENTED
 CHIRAL CONDENSATES }
\maketitle
\begin{center}
{\it{\bf D. Boyanovsky$^{(a)}$, H.J. de Vega$^{(b)}$, 
R. Holman$^{(c)}$, S. Prem Kumar$^{(c)}$}} \\
{\it{(a) Department of Physics and Astronomy, University of Pittsburgh,
PA. 15260,\\U.S.A.}} \\
{\it{(b) LPTHE, Universite' Pierre et Marie Curie (Paris VI), Tour 16,
1er. etage,\\4, Place Jussieu 75252 Paris, Cedex 05, France}}\\
{\it{(c) Department of Physics, Carnegie Mellon
University, Pittsburgh,PA. 15213,\\U.S.A.}}
\end{center}
\begin{abstract}
We study photoproduction during the non-equilibrium stages of the formation of
chiral condensates within the ``quench'' scenario of the chiral phase
transition. The dynamics is modeled with a gauged linear sigma model. A novel
quantum kinetic approach to the description of photoproduction far off
equilibrium is developed.  We find that non-equilibrium spinodal instabilities
of long wavelength pion fluctuations are responsible for an enhanced
photoproduction rate for energies $\leq 80$ MeV at order $\alpha$. These
non-equilibrium effects lead to a larger contribution than the typical
processes in the medium, including that of the anomalous neutral pion decay
$\pi^0 \rightarrow 2 \gamma$ (which is of order $\alpha^2$).  We follow the
evolution of the dynamics throughout the phase transition, which in this
scenario occurs on a time scale of about $2.5-3$ fm/c and integrate the photon
yield through its evolution.  The spectrum of photons produced throughout the
phase transition is a non- equilibrium one. For thermal initial conditions at
the time of the quench it interpolates between a thermal distribution about
$6\%$ above the initial temperature (at the time of the quench) for low energy
$\leq 80$ MeV photons, and a high energy tail in thermal equilibrium at the
initial temperature, with a smooth crossover at 100 MeV. The rate displays a
peak at $\sim 35$ MeV which receives a larger enhancement the closer the
initial temperature at the time of the quench is to the critical
temperature. It is found that the enhancement of photoproduction at low
energies is not an artifact caused by the initial distribution of the photons,
but is due to the pionic instabilities. We suggest that these strong out of
equilibrium effects may provide experimental signatures for the formation and
relaxation of DCC's in heavy ion collisions. 
\end{abstract}
\section{Introduction}
Considerable interest has been sparked recently by the possibility that
disoriented chiral condensates (DCC's) could form during the early stages of
evolution in high-energy, high luminosity hadron or heavy-ion collisions since
this would be a signal of the chiral phase
transition\cite{anselm1,blaizot,bjorken1,kowalski,bjorken2,revs}. The main idea
is that near the central rapidity region in these collisions, typically a large
amount of energy $\sim$ a few Gev is deposited in a volume of order a fm$^3$,
corresponding to temperatures above $200$ MeV at which the chiral symmetry is
restored\cite{bjor,book1,book2}. It is conceivable that as these regions cool
down through hydrodynamic expansion they might give rise to domains where the
order parameter of QCD points in a different direction from the $\sigma$ one,
resulting in a disoriented chiral condensate
(DCC's)\cite{anselm1,blaizot,bjorken1,kowalski,bjorken2,revs}. It has been
speculated that these regions will then act as strong pion lasers, relaxing to
the true physical vacuum on time scales of a few fm/c by emitting coherent
bursts of pions with very definite isospin
correlations\cite{bjorken1,kowalski,bjorken2}. This phenomenon would be a
striking experimental signature of the chiral phase transition and could
provide an explanation for the Centauro and anti-Centauro (JACEE) cosmic ray
events \cite{bjorken2,revs,anselm2,lattes}.

Since trying to understand such phenomena by investigating QCD is currently
impossible, efforts have mainly been focussed on the dynamics of low energy
effective theories of hadrons, some popular examples being the linear and
non-linear $\sigma$ models.  The rationale for using a low energy effective
theory approach is that, as in the study of critical phenomena, one expects
that if a large amplitude coherent configuration of soft modes is produced
during the transition, such an occurence should be {\em universal} and only
very weakly dependent on the detailed features of the theory.

Rajagopal and Wilczek\cite{wilraj1} have argued that the $O(4)$ linear $\sigma$
model encodes the relevant low energy chiral phenomenology of QCD.  In
particular they argue that the model lies in the same {\em static} universality
class as QCD with 2 light flavours, and that \cite{wilraj2} large coherent
regions of correlated pions can form {\em only} if the phase transition is
strongly out of equilibrium. They have proposed the ``quench'' scenario in
which the initial ``hot region'' supercools through the phase transition via
hydrodynamical expansion on time scales much shorter than the relaxation times
for long-wavelength fluctuations. As a result, the system enters into the
spinodal region which induces the growth of the unstable long-wavelength modes
just as in the process of spinodal decomposition in phase
separation\cite{boysinlee}. This version of a disoriented chiral condensate is
somewhat different from the ``baked alaska'' scenario envisaged by
Bjorken,Kowalski and Taylor\cite{bjorken1,kowalski,bjorken2} although the
relaxation process of both configurations is similar.

Results obtained from studies of both the classical
\cite{wilraj2,gavin,randrup,huangwang} and quantum linear sigma model
\cite{cooper,boyan}, including hydrodynamical expansion\cite{cooper} seem to
suggest that there is a range of initial conditions for which there is
amplification of pion fluctuations that could lead to the formation of large
coherent domains during the chiral phase transition. However, pions are
strongly interacting hadrons, with a typical mean free path in medium of about
2-3 fm \cite{goity} under conditions to be achieved at RHIC or LHC. This
results in strong final state interactions, and it could be that a DCC signal
may be indistinguishable from the background. 

There are currently two experiments searching for a DCC signal, T-864
test/experiment at the Fermilab Tevatron collider\cite{minimax} and WA98 at
CERN-SPS\cite{wa98} and a proposal for a full acceptance detector at LHC
(FELIX)\cite{wa98}. It is thus imperative to identify an alternative signature
of the chiral phase transition capable of providing information on the early
time evolution of the collision and the formation of DCC's or any other out of
equilibrium effect associated with the transition.  Photons and dileptons are
the most promising probes of the early stages of the transition. Since they
interact mainly electromagnetically, with mean free paths typically much larger
than the size of the fireball they carry undistorted information of the initial
non-equilibrium stages of the evolution of the plasma.

In this context, it is important to realize that there are many sources of
photons in relativistic heavy-ion collisions: (i) direct QCD (hard) photons
originating from hard partonic processes (typically Drell-Yan, and $q\bar{q}
\rightarrow g \gamma$), (ii) thermal and non-thermal photons originating from
the quark-gluon plasma, (iii) photons originating from the hadronic
gas\cite{alam}. It is the last process that is of interest to us, and is
usually the dominant process in the production of long wavelength photons with
energies less than a few hundreds of MeV's. Some of the main sources of photons
in this energy range are the decay of the neutral pion $\pi^0 \rightarrow
\gamma \gamma$, $\eta$ and $\eta'$ and resonance decay into photons.  Of these,
the leading decay channel for low energy photons is the ``anomalous'' decay of
the neutral pion.

In this article we concentrate solely on photo-production, with the particular
goal of obtaining an {\em ab initio}, real-time expression for the
photoproduction rate in {\em strongly} out of equilibrium situations.  Our main
premise is that {\em if} the chiral phase transition is ``quenched'' as
envisaged in the scenario of Rajagopal and Wilczek, with the long-wavelength
pion modes becoming unstable, they will grow almost exponentially in the
initial stages of the transition, so that these strongly out of equilibrium
processes {\em must} result in enhanced photoproduction at low energies.

We note that although we focus on the ``quench'' scenario within a specific
phenomenological model, the new kinetic approach developed in this article is
more general and not tied to this scenario. In particular, if the transition is
strongly first order and is driven by nucleation and growth of hadronic
bubbles, as suggested by Kapusta and Vischer\cite{kapvis}, this non-equilibrium
process should also lead to enhanced photoproduction.

Dilepton and photon production from a {\em classical} coherent pion source have
been studied by Huang and Wang\cite{wang}. These authors studied
photoproduction due to a classical Blaizot-Krzywicki configuration and found a
strong dependence on the initial conditions. However, to our knowledge there
has not yet been an attempt to compute the photoproduction rate including
quantum and thermal effects and strong non-equilibrium effects as should result
from the fast relaxation of DCC's. It is the point of this article that such a
study requires a novel formulation of the kinetics of photoproduction (and
eventually of dilepton production).


Typical calculations of photoproduction rates assume a system which is in
equilibrium or quasi-equilibrium (steady state), and show that the rates are
related to the Fourier transform of the hadronic current-current correlator
\cite{fein,toimela,rusk,kapustagale,lichard,zahed}. 



The main idea behind our approach is the calculation of expectation values of
time evolving observables such as the photon number from first principles,
using real-time non-equilibrium quantum field
theory\cite{schwinger,chou,kapusta} treating the pion instabilities
non-perturbatively\cite{boyan2}.  These non-perturbative methods were
introduced within the context of DCC formation in
references\cite{cooper,boyan}. Such framework allows us to go beyond the
standard results for the emission rates to study strongly out of equilibrium
situations.  In particular we find that the emission rates are given by the
hadronic (pion) current-current correlation functions, but these are obtained
from the full non-equilibrium dynamics of the pion fields. Thus the emission
rates and consequently the photon spectrum obtained using this approach reflect
all the long range correlations that were generated during the phase
transition. Moreover, the rate equation resembles a {\bf generalized kinetic
equation} that encodes all the non-equilibrium time scales of DCC formation,
growth and relaxation.  In particular, we notice the absence of energy
conservation at short time scales $\sim$ 1 fm/c, and hence photoproduction
proceeds via processes that would be kinematically forbidden in an equilibrium
situation.

The focus of our article is to study the photoproduction rates from the
coherent DCC regions which are formed during the {\em non-equilibrium} stages
during a `quenched' chiral phase transition including the quantum mechanical
and thermal effects.

In section II we give a summary of the non-equilibrium dynamics of the linear
sigma model. We do not incorporate expansion at this stage, and treat the
dynamics of the transition via a ``quench''. In section III we ``gauge'' the
sigma model and discuss its range of validity. In section IV, we present the
details of our novel approach to the kinetics of photoproduction as an initial
condition problem.  We obtain the remarkable result that in a situation far
from equilibrium, as is the case under consideration, the {\em rate} is
${\cal{O}}(\alpha)$ (with the full quantum mechanical current-current
correlation function). In section V we discuss the numerical results of the
calculations of photoproduction, and show that the distribution function of
produced photons is out of equilibrium interpolating between two local
equilibrium situations at different temperatures, with an enhanced
photoproduction rate for energies below 80 MeV. We also compare the results to
the case in which there are no photons initially and conclude that the
non-equilibrium enhancement has its origin solely on the pionic instabilities
and is insensitive to the initial photon distribution. The resulting photon
spectrum is out of equilibrium and non-thermal. We compare our results to those
obtained recently on photoproduction from the hadronic gas\cite{lichard,zahed}. Such a
comparison unequivocally reveals the non-equilibrium features associated with
the dynamics of the phase transition.  In Section VI we discuss the
contribution from $\pi^0 \rightarrow 2 \gamma$ out of equilibrium and argue
that it is {\em subleading} compared to the lowest order contribution from the
charged pions. In section VII we present our conclusions, summarize the main
results and pose new questions. Two appendices are devoted to technical details
and a pedagogical example of the novel kinetic approach.
     
\section{The O(4)linear sigma model out of equilibrium}

The linear sigma model has been one of the most popular of the phenomenological
models to describe the chiral phase transition because it incorporates the
correct symmetries and current algebra relations.  Obviously such a simple
description cannot capture all of the details of the underlying theory, QCD,
but if the chiral phase transition entails the formation of long-wavelength,
coherent pion clouds, one expects this to be a robust feature, fairly
independent of the detailed confining dynamics.

This model has been studied intensely from the classical,
\cite{wilraj1,wilraj2,gavin,huangwang} semiclassical \cite{randrup}, and
quantum perspectives both with \cite{cooper} and without \cite{boyan}
expansion.  In this article we do not incorporate either longitudinal or
spherical expansion \cite{randrup,cooper}, but prefer to illustrate the main
concepts in the simplest of settings, that of a ``quenched'' transition. If (as
we will see later) interesting new physics emerges, there would be motivation
to include expansion and other features such as inhomogeneities in the analysis
of photoproduction from the dynamics of the Disoriented Chiral
Condensates. Here we summarize the main features of the dynamics in the
``quench'' scenario of the {\em ungauged} linear sigma model, and postpone
until the next section the subtle details of its gauging.
 
The Lagrangian density for the linear sigma model with an explicit symmetry
breaking term is given by

\begin{equation}
{\cal{L}} = \frac{1}{2}\partial_{\mu}\vec{\Phi}\cdot
\partial^{\mu}\vec{\Phi}-\frac{1}{2}m^2(t)
\vec{\Phi}\cdot\vec{\Phi}+\lambda 
(\vec{\Phi}\cdot\vec{\Phi})^2-h\sigma \label{lagrangian}
\end{equation}
where $\vec{\Phi}$ is an $O(N+1)$ vector, $\vec{\Phi}=(\sigma,\vec{\pi})$ and
$\vec{\pi}$ represents the $N$ pions,
$\vec{\pi}=(\pi^0,\pi^1,\pi^2,\pi^3,....\pi^{N-1})$, and $m^2(t)$ introduces
phenomenologically ``by hand'' a quench situation. Comparison between the
``quench'' approximation\cite{boyan} and more realistic treatments including
longitudinal and spherical hydrodynamical expansions
reveal\cite{cooper,randrup} that there is no great discrepancy between the
different results insofar as the time scales and correlations.

We then take

\begin{equation}
m^2(t)= \frac{M^2_{\sigma}}{2}\left[\frac{T^2_i}{T^2_c}\Theta(-t)-1\right]
\; \; ; T_i \geq T_c
\label{massoft1}
\end{equation} 
corresponding to a ``quench'' from the high temperature phase above the
critical temperature, to zero temperature.

Although we will be dealing with the $O(4)$ model with 3 pions, we will use the
large $N$ limit in order to provide a consistent, non-perturbative framework to
study the dynamics.  It can be implemented by use of an auxiliary field
consistently to all orders in $1/N$\cite{cooper2}. This approximation is
renormalizable, energy conserving and maintains all of the Ward identities
associated with chiral symmetry and current algebra.  


As a result, the pions in the asymptotic equilibrium state are exactly massless
in the absence of the explicit symmetry breaking term. A consequence of
Goldstone's theorem is that the spinodal line {\em coincides} with the
coexistence line, so that there are long-wavelength instabilities
for all values of the order parameter between zero and the equilibrium minimum
$f_{\pi}$.
  
The non-equilibrium equations of motion are obtained via the tadpole
method\cite{boyan2}: first shift $\sigma$ by its expectation value in the
non-equilibrium state
\begin{equation}
\sigma(\vec{x},t)=\phi(t)+\chi(\vec{x},t) \; \; ; \;  \phi(t)=
\langle \sigma(\vec x,t) \rangle \label{split}
\end{equation}
and then implement the tadpole condition
\begin{equation}
\langle\chi(\vec{x},t)\rangle=0 \; \; ; \; 
\langle\vec{\pi}(\vec{x},t)\rangle=0 \label{tadpole}
\end{equation}
to all orders in the corresponding expansion. The expectation values of the
Heisenberg operators $\chi(\vec{x},t)$ and $\vec{\pi}(\vec{x},t)$ above are to
be computed in the initial density matrix, or alternatively, we can compute
these expectation values using the Schrodinger picture operators and the
time-evolved density matrix, as described below.

To leading order in $1/N$ the auxiliary field method\cite{cooper2} is
equivalent to the following Hartree like factorization\cite{boyan2}:
\begin{equation}
\chi^4 \rightarrow 6\langle\chi^2\rangle\chi^2+\text{constant} 
\; \; ; \; 
\chi^3 \rightarrow 3\langle\chi^2\rangle\chi \label{factor1}
\end{equation}
\begin{equation}
(\vec{\pi}\cdot\vec{\pi})^2 \rightarrow (2+\frac{4}{N}) 
\langle{\vec{\pi}}^2\rangle{\vec{\pi}}^2+\text{constant}
\label{factor2}
\end{equation}
\begin{equation}
{\vec{\pi}}^2\chi^2 \rightarrow {\vec{\pi}}^2\langle\chi^2\rangle+ 
\langle{\vec{\pi}}^2\rangle\chi^2 \; \; ; \; \;  {\vec{\pi}}^2\chi \rightarrow
\langle{\vec{\pi}}^2\rangle\chi.  
\label{factor3}
\end{equation}

Non-equilibrium quantum field theory requires a path integral representation
along a complex countour in time\cite{schwinger,chou}, with the Lagrangian
density along this contour given by
\begin{equation}
{\cal{L}}_{neq} = {\cal{L}}[\Phi^+]-{\cal{L}}[\Phi^-]
\end{equation}
with the fields $\Phi^{\pm}$ defined along the forward ($+$) and backward ($-$)
time branches. For further details see references\cite{cooper2,boyan2}.  In
the leading order in $1/N$, the effective non-equilibrium Lagrangian density is
given by

\begin{eqnarray}
&&{\cal{L}}[\phi+\chi^+,\vec{\pi}^+]-{\cal{L}}[\phi+\chi^-,
 \vec{\pi}^-]= \left\{\frac{1}{2}(\partial\chi^+)^2+\frac{1}{2}
(\partial\vec{\pi}^+)^2 \right. \nonumber \\
&&\left. -{\cal{V}}^{\prime}(t)\chi^+- \frac{1}{2}{\cal{M}_{\chi}}^{+2}(t)
{\chi}^{+2}-\frac{1}{2}{\cal{M}_{\vec{\pi}}}^{+2}(t){\vec{\pi}}^{+2}\right\}
-\left\{+\longrightarrow-\right\}.\label{leff}
\end{eqnarray}
where
\begin{equation}
{\cal{V}}^{\prime}(t)=
\ddot{\phi}(t)+\phi(t)[m^2(t)+4\lambda\phi^2 (t) + 4 \lambda
\langle{\vec{\pi}}^2 \rangle(t)]-h \label{vprime}
\end{equation}
\begin{equation}
{\cal{M}}_{\vec{\pi}}^2(t) = m^2(t)+4\lambda\phi^2 (t) + 4 \lambda
\langle{\vec{\pi}}^2 \rangle(t)\label{pionmass}
\end{equation}
\begin{equation}
{\cal{M}}_{\chi}^2(t) = m^2(t)+12\lambda\phi^2 (t) + 4 \lambda
\langle{\vec{\pi}}^2 \rangle(t).\label{chimass}
\end{equation}

Comparing eqns.(\ref{vprime}, \ref{pionmass}) when $\ddot{\phi} = 0$ (the
equilibrium case) one finds the PCAC relation
\begin{equation}
f_{\pi}M^2_{\pi}= h \label{pcac}
\end{equation}
with $f_{\pi}$ the equilibrium expectation value of the sigma field and
$M_{\pi}$ the equilibrium pion mass. The validity of this relation in this
approximation is a consequence of the large $N$ expansion. A naive Hartree
factorization would violate this fundamental Ward identity.  To leading order
in the $1/N$ expansion, the theory is quadratic at the expense of a
self-consistent condition and the fluctuations of the sigma field do not
influence the dynamics of the expectation value or the pion fields and thus
decouple from the dynamics. The pion fields obey a linear Heisenberg evolution
equation (in terms of the self-consistent field), and the Heisenberg field
operators can be expanded as

\begin{equation}
\vec{\pi}(\vec{x},t)=\frac{1}{\sqrt{\Omega}}\sum_k\frac{1}{\sqrt{2W_{k,i}}}
[\vec{a}_kU_k(t)e^{i\vec{k}\cdot\vec{x}}+\vec{a}^{\dagger}_kU_k^*(t)e^{-i\vec{k}\cdot
\vec{x}}]
\end{equation}
where $a_k$, $a_k^\dagger$ are the destruction and creation operators of Fock
states associated with the pion field, $\Omega$ is the quantization volume, and
the mode functions $U_k(t)$ and the order parameter $\phi(t)$ obey:
\begin{equation}
\ddot{\phi}(t)+[m^2(t)+4\lambda\phi^2(t)+4\lambda\langle{\vec{\pi}^2}
\rangle(t)] \phi(t)-h=0 \; \; ; \; \; \phi(0)=0 ; \dot{\phi}(0)=0
 \label{zeromodeeqn}
\end{equation}
\begin{eqnarray}
&&\left[\frac{d^2}{dt^2}
+k^2+m^2(t)+4\lambda\phi^2(t)+4\lambda\langle\vec{\pi}^2
\rangle(t)\right]U_k(t)=0\label{modeqns}
\\
&&\hspace{0.1in}U_k(0)=1;\hspace{0.1in}\dot{U}_k(0)
=-iW_{k,i} \; \; ; \; \; W_{k,i}=\sqrt{k^2+m^2(t<0)} \label{inicon}
\end{eqnarray}
\begin{equation}
\langle\vec{\pi}^2
\rangle(t)=N\int\frac{d^3k}{(2\pi)^3}
\frac{|U_k(t)|^2-1}{2W_{k,i}}\coth\left
[\frac {W_{k,i}}{2T_i}\right]\label{fluct}
\end{equation} 

Our initial conditions on the pion modes are that at the time of the ``quench'',
the pion modes were in local thermodynamic equilibrium at the initial
temperature $T_i$ with the effective mass $m(t<0)$ given by
(\ref{massoft1}). The initial condition on $\phi$ is chosen so that the
expectation value of the sigma field was at the ``top'' of the potential hill,
and the self-consistent fluctuation has been subtracted at the initial time ,
thus renormalizing the mass (for more details see reference\cite{boyan}).

The linear sigma model is a phenomenological model, and its parameters are
fixed by the low energy pion physics to be:

\begin{eqnarray}
	M_{\sigma} \approx 600 \mbox{ MeV }     & & \; \; ; \; \;
		f_{\pi} \approx
95 \mbox{ MeV } \; \; ; \; \; \lambda \approx 4.5 \nonumber \\
h \approx (120 \mbox{ MeV })^3             & & \; \; ; \; \; 
	T_c \approx 200 \mbox{ MeV } .
\label{parameters}
\end{eqnarray}

The critical temperature $T_c$ is a consequence of the parameters of the model
and is somewhat larger than the lattice estimates $T_c \approx 150$ MeV. The
sigma model must be treated as a cutoff theory with an ultraviolet cutoff of
the order of $\Lambda \approx 1-2$ GeV. There are two reasons for this
cutoff: i) for such a large value of the coupling constant the Landau pole is
at the order of this scale if the parameters (the coupling) are determined at
the scale $\approx M_{\sigma}$, ii) more fundamentally, the simple linear sigma
model does not incorporate higher mass hadrons, such as vector meson resonances
(with masses $\geq 770 \mbox{Mev}$) and nucleons.  The time evolution will only
be sensitive to the scale of this cutoff for time scales much smaller than
about $0.1 \ \mbox{fm}/c$, but we are interested in longer time scales of a few
fm/c and the numerical results on these time scales proved to be insensitive to
the values of the cutoff in the range 1-2 GeV.

The self-consistent numerical solution to the coupled non-linear equations
(\ref{zeromodeeqn},\ref{modeqns},\ref{inicon},\ref{fluct}) with the above
values of the parameters leads to a complete description of the dynamics. The
non-equilibrium Green's functions for the pion fields, which will be necessary
in the calculation of photoproduction, are completely determined by the mode
functions $U_k(t)$ which are solutions to the set of self-consistent equations
(\ref{modeqns},\ref{fluct}).  In particular, the spatial Fourier transform of
the necessary Green's functions are given by :

\begin{equation}
{\langle}{\pi}^{+}_a(\vec{k},t){\pi}^{+}_b(-{\vec{k}},
t^{\prime}){\rangle}=-i\delta_{a,b}\left[
G_k^{>}(t;t^{\prime})
\Theta(t-t^{\prime})+G_k^{<}(t;t^{\prime})
\Theta(t^{\prime}-t) \right] \label{gplpl}
\end{equation}
\begin{equation}
{\langle}{\pi}^{-}_a(\vec{k},t){\pi}^{-}_b
(-{\vec{k}},t^{\prime}){\rangle}= -i \delta_{a,b}\left[
G_k^{>}(t;t^{\prime})
\Theta(t^{\prime}-t)+G_k^{<}(t;t^{\prime})
\Theta(t-t^{\prime}) \right] \label{glele}
\end{equation}
\begin{equation}
{\langle}{\pi}^{+(-)}_a(\vec{k},t){\pi}^{-(+)}_b
(-{\vec{k}},t^{\prime}){\rangle}= -i\delta_{a,b}
G_k^{<(>)}(t;t^{\prime})
\label{gplmin}
\end{equation}
As a consequence of assuming that the system was initially in local
thermodynamic equilibrium at the temperature $T_i = 1/\beta$, the Green's
functions $G^< , G^>$ obey the KMS condition
\begin{equation}
G_k^<(t_0;t)= G_k^>(t_0-i\beta;t).
\end{equation}
The fundamental Wightman functions $G^{<(>)}$ are constructed from the pion
mode functions, solutions of (\ref{modeqns}) as:
\begin{equation}
G_k^>(t,t^{\prime})=\frac{i}{2W_{k,i}}[(1+n_b)U_k(t)U_k^*(t^\prime)
+n_bU_k^*(t)U_k(t^\prime)]\label{gfourier>}
\end{equation}
\begin{eqnarray}
&& G_k^<(t,t^{\prime})=\frac{i}{2W_k}[(1+n_b)U_k^*(t)U_k(t^\prime)
+n_bU_k(t)U_k^*(t^\prime)]\label{gfourier<}\\
&& n_b=\frac{1}{e^{\beta W_{k,i}}-1} \label{ocunum}
\end{eqnarray}

The numerical results of the integration of the self-consistent evolution
equations given above have already been discussed in the
literature\cite{boyan}. The main feature of the dynamics is that for early
times $t>0$ a band of long wavevectors ``see'' an inverted harmonic oscillator
and grow almost exponentially until the backreaction of the fluctuations and
the evolution of the order parameter shut these instabilities off. These are
the ``spinodal'' instabilities that drive the phase transition out of
equilibrium.  The numerical analysis reveals that at early times long-
wavelength pion modes with wavevectors $k \leq 200$ MeV grow almost
exponentially, whereas the zero mode of the sigma field, i.e.  the order
parameter, rolls towards its equilibrium value close to $f_{\pi}$ on a time
scale $\approx 3 \mbox{ fm}/c$. Figure 1 shows the evolution of $f(\tau)=
\phi(t)/M_F$ vs $\tau = cM_F t$ ($M_F=200\ \mbox{MeV}= 1/ \mbox{fm}$). As shown
in figure 1, at time scales of the order of $3 \mbox{ fm}/c$ the order
parameter reaches $f_{\pi}$ and begins to oscillate around this value,
indicating that the phase transition is almost complete.  Parametrizing
$U_k(t)= |U_k(t)|e^{-i\varphi_k(t)}$, we show in figure 2.a $\ln(|U_k(t)|)$ vs
$\tau$ (time in units of fm/c) for several values of $k$. Figure 2.b shows the
values of $\varphi_k(t)$ vs $\tau$ for the same values of $k$ as in figure 2.a.
Clearly there is a band of long-wavelength pion modes $k \leq 150-200$ MeV that
grow almost exponentially during the time for which the order parameter rolls
down the potential hill towards $f_{\pi}$ and their phases vary smoothly.  The
reason that we focus on the phases will become clear when we derive the kinetic
equations for photoproduction.  In short, since the phases vary rather smoothly
and monotonically during this interval, there is no large phase cancellation
due to ``dephasing'' of these modes so that they will contribute coherently to
photoproduction. These long wavelengths unstable modes do not ``decohere''
during the relevant time scales and as we will see when we study the kinetics, they
give the primary contribution to the non-equilibrium photoproduction rate.

\section{Gauging the linear sigma model}

Both charged and neutral pions couple to electromagnetism. Whereas the
charged pions couple through the charge form factor, the neutral pions
couple through the chiral anomaly. In the vector meson dominance 
approximation, the charge form factor of the (charged) pion is 
determined by the exchange of a $\rho$ vector meson and given 
by\cite{donoghue}
\begin{equation}
G_{\pi}(q^2) \approx 1+ \frac{q^2}{m_{\rho}^2}+\cdots
\label{pionformfactor}
\end{equation}
with $m_{\rho} \approx 770$ MeV is the mass of the $\rho$ vector
meson. 

Since we will be focusing on on-shell photoproduction of low momentum photons
we can approximate the charge form factor by 1 to the order that we are
calculating (no off-shell photons in intermediate states).

The coupling of the neutral pion is determined by the anomalous quark triangle
diagram, which for three colors and two quark flavors (up and down) gives the
effective ``anomalous'' vertex
\begin{equation}
{\cal{L}}_A = \frac{\alpha}{8 \pi f_{\pi}}
\epsilon^{\mu \nu \alpha \beta}F_{\mu \nu}F_{\alpha \beta} \pi^0
\label{anomalousvertex}
\end{equation}

The physical charged pion fields are given by the combinations

\begin{equation}
\pi_+=\frac{\pi^1+i\pi^2}{\sqrt{2}},\hspace{.5in}\pi_-=\frac{\pi^1-i\pi^2}
{\sqrt{2}}.\label{picomplex}
\end{equation}
whereas the neutral pion is identified with the third isospin component $\pi^0$
and the effective lagrangian with electromagnetic coupling in this
approximation is given by
\begin{equation}
{\cal{L}} = \frac{1}{2}(\partial_{\mu}\sigma)^2 + \frac{1}{2}
(\partial_{\mu}\vec{\pi})^2 + \frac{e^2}{2} (\pi_1^{2}+
\pi_2^{2})A_{\mu}A^{\mu} + j_{\mu}A^{\mu}-
\frac{1}{4}F_{\mu \nu}F^{\mu \nu}+ {\cal{L}}_A -V(\sigma,\vec{\pi})
\label{veff}
\end{equation}
\begin{equation}
j_{\mu}= e(\pi_1 \partial_{\mu}\pi_2-\pi_2\partial_{\mu}\pi_1)
\label{electrocurrent}
\end{equation}
with $V(\sigma,\pi)$ is given by the potential term in
(\ref{lagrangian}) in terms of $\sigma , \pi_{1,2}, \pi^0$. The 
current $j_{\mu}$ is identified with the third component of the
isospin current.

After performing the shift $\sigma \rightarrow \phi(t)+\chi(\vec x,t)$,
the total effective non-equilibrium Lagrangian is given by
\begin{eqnarray}
{\cal{L}}_{neq}  & = &
{\cal{L}}[\phi+\chi^+,\vec{\pi}^+,A_\mu^+]-{\cal{L}}[\phi+\chi^-,\vec{\pi}^-,
A_\mu^-]=\frac{1}{2}(\partial\chi^+)^2+\frac{1}{2}
(\partial\vec{\pi}^+)^2 
\\\nonumber
&-&\chi^+\ddot{\phi}-\frac{1}{4}F^{+\mu\nu}F^{+}_{\mu\nu}+
\frac{e^2}{2}(\pi_1^{+2}+\pi_2^{+2})A_{\mu}^
+ A^{+\mu}+j_{\mu}^+A^{+\mu}+{\cal{L}}_A(\pi_3^+,A_{\mu}^+)\\\nonumber
&-&\left[V^{\prime}(\sigma,\vec{\pi}^+)\chi^++\frac{1}{2!}
V^{\prime\prime}(\sigma,\vec{\pi}^+)\chi^{+2}+\frac{1}{3!}V^{[3]}
(\sigma,\vec{\pi}^+) \chi^{+3}+\frac{1}{4!}V^{[4]}
(\sigma,\vec{\pi}^+) \chi^{+4}\right]\\\nonumber
&-&(+\longrightarrow-).
\\\nonumber
\end{eqnarray}
The fields $\vec{\pi}^+$ should not be confused with the charged pion $\pi^+$;
the superscript in the above expression refers to the time branches. However
before performing calculations with the above non-equilibrium Lagrangian we
must address the issue of gauge invariance and determine the physical
observables. 

Since we want to avoid potential ambiguities with gauge fixing terms and gauge
choices, we will describe our quantization procedure solely in terms of
physical degrees of freedom. This is best achieved by passing to the
Hamiltonian formulation, using it to recognize the physical degrees of freedom
and then casting the non-equilibrium calculation solely in terms of these, thus
avoiding any potential ambiguity with gauge artifacts.

\subsection{The Electromagnetic Sector:}

Since we are dealing with an Abelian gauge theory, there are only two first
class constraints for the gauge sector: vanishing canonical momentum for $A_0$
and Gauss's law (the generator of time independent gauge transformations). We
can work in terms of gauge invariant observables \cite{boygauge} by projecting
on the states in the Hilbert space that are annihilated by these first class
constraints, or alternatively fix Coulomb gauge which is a physical gauge
displaying the two transverse physical photon polarizations and the
instantaneous Coulomb interaction.

Our first goal is to express the number operator for asymptotic, physical
photons with two massless transverse degrees of freedom, in a convenient form
that can be inserted into the non-equilibrium path integral.  In a plasma the
``longitudinal'' component (the instantaneous Coulomb interaction) is screened
by a Debye screening length or electric mass, but the transverse components are
not screened as there is no magnetic mass in the Abelian theory.
 
We concentrate only on the physical components i.e. the transverse components
of the photon field $\vec{A}_T(\vec{x},t)$. If $\vec{\Pi}_T$ represents the
momentum conjugate to $\vec{A}_T(\vec{x},t)$, we can write the Hamiltonian for
the free electromagnetic theory as follows:
\begin{equation}
H=\int d^3x \left[\frac{1}{2}\vec{\Pi}_T\cdot\vec{\Pi}_T+ \frac{1} {2} ( \vec
{\nabla}\times \vec{A}_T)^2\right]. \label{freegauge}
\end{equation} 
We then define the Fourier components of the fields as
\begin{equation}
\vec{P}_{T}(\vec{k})=\int \frac{d^3x}{\sqrt{\Omega}}\vec{\Pi}_T(\vec x)
e^{-i\vec{k}\cdot\vec{x}}\label{pitrans} 
\end{equation}
\begin{equation}
\vec{\Phi}_{T}(\vec{k})=\int \frac{d^3x}{\sqrt{\Omega}}\vec{A}_T(\vec x)
e^{-i\vec{k}\cdot\vec{x}},\label{phitrans}
\end{equation}
where $\Omega$ is the quantization volume. In terms of these variables the
Hamiltonian becomes
\begin{equation}
H=\frac{1}{2}
\sum_{\vec{k}}\left[\vec{P}_{T}(\vec{k})\cdot\vec{P}_{T}(-\vec{k})
+k^2\vec{\Phi}_{T}(\vec{k})\cdot\vec{\Phi}_{T}(-\vec{k})\right].
\end{equation}
Thus the number operator for photons with momentum $\vec{k}$ {\em per
polarization} is given by the average over the two polarizations
\begin{equation}
N_{\vec{k}}=\frac{1}{4k}\left[\vec{P}_{T}(\vec{k})
\cdot\vec{P}_{T}(-\vec{k})
+k^2\vec{\Phi}_{T}(\vec{k})\cdot\vec{\Phi}_{T}(-\vec{k})\right]-\frac{1}{2}.
\label{number1}  
\end{equation}
This expression, which is more amenable for use in the non-equilibrium
formulation is equivalent to the familiar one given by
\begin{equation}
N_{\vec{k}}=\frac{1}{2}\sum_{\lambda}N_{\vec{k}(\lambda)}
=\frac{1}{4}\sum_{\lambda}
\left[b^{\dagger}_{\vec{k}(\lambda)}b_{\vec{k}(\lambda)}
+b^{\dagger}_{-\vec{k}(\lambda)}b_{-\vec{k}(\lambda)}\right].
\label{number2}
\end{equation}
which gives the average number of transverse photons {\em per polarization}.
Here $\lambda$ indicates the transverse polarization states, while
$b^{\dagger}_{\vec{k}(\lambda)}$ and $b_{\vec{k}(\lambda)}$ are the creation
and destruction operators for photons of momentum $\vec{k}$ and polarization
$\lambda$. The equivalence is made manifest through the mode expansions:
\begin{equation}
\vec{A}_T(\vec{x},t)=\frac{1}{\sqrt{\Omega}}\sum_{\vec{k},\lambda}
\frac{\vec{\epsilon} _{\lambda} (\vec{k} )}{\sqrt{2k}}
\left[ b_{\vec{k}(\lambda)} e^{i\vec{k}\cdot\vec{x}} 
+b^{\dagger}_{\vec{k}(\lambda)} e^{-i\vec{k}\cdot\vec{x}}\right]
\label{modea}
\end{equation}
\begin{equation}
\vec{\Pi}_T(\vec{x},t)=\frac{-i}{\sqrt{\Omega}}\sum_{\vec{k},\lambda}
\vec{\epsilon} _{\lambda} (\vec{k})
{\sqrt{\frac{k}{2}}}
\left
[ b_{\vec{k}(\lambda)} e^{i\vec{k}\cdot\vec{x}} 
-b^{\dagger}_{\vec{k}(\lambda)} e^{-i\vec{k}\cdot\vec{x}}\right].
\label{modepi}
\end{equation}
Where the $\vec{\epsilon}$'s are the transverse polarization vectors.
Combining equations (\ref{pitrans}), (\ref{phitrans}), (\ref{modea}),
(\ref{modepi}), we obtain the Fourier components of the transverse phase space
variables to be
\begin{equation}
\vec{\Phi}_T(\vec{k})=\frac{1}{\sqrt{2k}}
\sum_{\lambda}\left[b_{\vec{k}(\lambda)}\vec{\epsilon} _{\lambda} (\vec{k})
+ b^{\dagger}_{-\vec{k}(\lambda)}\vec{\epsilon} _{\lambda} (-\vec{k})
\right]\\
\end{equation}
\begin{equation}
\vec{P}_T(\vec{k})=-i\sqrt{\frac{k}{2}}
\sum_{\lambda}\left[b_{\vec{k}(\lambda)}\vec{\epsilon} _{\lambda} (\vec{k})
-b^{\dagger}_{-\vec{k}(\lambda)}\vec{\epsilon} _{\lambda} (-\vec{k})
\right]
\end{equation}
These identifications lead at once to the equivalence between the number
operators given by (\ref{number1},\ref{number2}) up to a normal ordering
constant.

In order to obtain an equation of motion for this number operator we now must
address the interactions.
 
\subsection{The Charged Sector of the Sigma-Model}

In this section we will focus on the interaction between the charged pions and
the electromagnetic field, under the assumption of a local interaction (charge
form factor approximated by one). We will postpone the discussion of the
anomalous coupling of the neutral pion to a later section, where we will argue
that such a contribution is subleading out of equilibrium.  The Hamiltonian (in
the gauge invariant sector) for the charged sector is given by
\begin{eqnarray}
H=&&\int d^3x\left[\frac{1}{2}\vec{\Pi}_T\cdot\vec{\Pi}_T
+\frac{1}{2}(\vec{\nabla}\times\vec{A}_T)^2
+(\vec{\nabla}\pi^+)\cdot(\vec{\nabla}\pi^-)+e^2\vec{A}_T\cdot\vec{A}_T
\pi^+\pi^-\right.\\\nonumber
&&\left.+ie\vec{A}_T\cdot[\pi^+\vec{\nabla}\pi^--\pi^-\vec{\nabla}\pi^+]+
\text{Coulomb term}\right].
\label{hphys}
\end{eqnarray}

The Coulomb interaction will appear in the photoproduction rate at
${\cal{O}}(\alpha^2)$ but not to order ($\alpha$); we will neglect this term in
our lowest order calculation.  A Hartree factorization of the ``seagull term'',
consistent with the large $N$ approximation, leads to a time dependent ``mass''
term for the transverse components

\begin{eqnarray}
&&e^2\vec{A}_T\cdot\vec{A}_T\pi^+\pi^-=\frac{1}{2}(e^2{\vec{A}_T}^2
(\pi_1^2+\pi_2^2))
\rightarrow\frac{1}{2}(e^2{\vec{A}_T}^2\langle\pi_1^2+\pi_2^2\rangle)\\
&&=(e^2{\vec{A}_T}^2\langle\pi_1^2\rangle)=\mu^2(t){\vec{A}_T}^2
,\label{photomass} 
\end{eqnarray}

The effect of this time dependent ``mass term'' is to ``squeeze'' the quantum
state for the transverse photon states, contributing to
photoproduction. However, as it will be clear later, such a contribution is of
${\cal{O}}(\alpha^2)$ and thus higher order in the electromagnetic coupling as
compared to the lowest order process that will be seen to dominate the
non-equilibrium photoproduction rate.

The electromagnetic current can be rewritten in terms of the pion fields
$\pi_1$ and $\pi_2$ using (\ref{picomplex}):
\begin{equation}
\vec{j}(\vec{x},t)=e[\pi_2\vec{\nabla}\pi_1-\pi_1\vec{\nabla}\pi_2].
\end{equation}
In terms of the spatial Fourier transform of the pion and gauge fields, we
obtain the following interaction vertex in momentum space
\begin{equation}
\int d^3x\vec{j}.\vec{A}_T= \sum_{\vec{p}}\vec{J}(-\vec p) \cdot 
\vec{\Phi}_T(\vec p) \; \; ; \; \; \vec{J}(-\vec p)= 
\frac{2ie}{\Omega}\sum_{\vec{p},\vec{k}}
\tilde{\pi}_1(\vec{k},t)
\tilde{\pi}_2(-\vec{k}-\vec{p},t)(\vec{k}\cdot\vec{\Phi}_T(\vec{p},t)),
\end{equation}
where we have used the fact that $\vec{p}\cdot \vec{\Phi}_T(\vec{p},t)=0$. The
interaction Hamiltonian, including the time dependent mass term for the photon
arising from the large $N$ factorization of the seagull term, but neglecting
the Coulomb interaction is now written as
\begin{eqnarray}
&&H=\sum_{\vec{q}}\left[\frac{1}{2}\vec{P}_{T}(\vec{q})\cdot\vec{P}_{T}(-\vec{q})
+\frac{1}{2}\omega_{\vec{q}}^2(t)\vec{\Phi}_{T}(\vec{q})\cdot\vec{\Phi}_{T}
(-\vec{q})+\vec{J}(-\vec{q})\cdot\vec{\Phi}_T(\vec{q})\right]\\
&&\omega_{\vec{q}}^2(t)=q^2+\mu^2(t),
\end{eqnarray}
leading to the Heisenberg equations of motion:
\begin{eqnarray}
&&\dot{\vec{\Phi}}_T(\vec{q},t)=\vec{P}_T(\vec{q},t)\\
&&\dot{\vec{P}}_T(\vec{q},t)=-\omega_{\vec{q}}^2(t)\vec{\Phi}_T(\vec{q},t)
-\vec{J}_T(\vec{q},t).
\end{eqnarray}
Here, $\vec{J}_T$ denotes the transverse component of the current, obtained by
projecting the full current onto the transverse polarization states:
\begin{equation}
J_{Ti}(\vec k)={\cal{P}}_{il}(\vec k)J_{l}(\vec k) \; \; ; \; \; 
{\cal{P}}_{il}(\vec k)=\delta_{il}-\frac{k_ik_l}{k^2}. \label{polarization}
\end{equation}
Using the definition of the number operator (\ref{number1}) and the Heisenberg
equations of motion, we obtain
\begin{equation}
\dot{N}_{\vec{q}}(t)= 
-\frac{1}{4q}\left[\mu^2(t)(\vec{\Phi}_T(\vec{q})\cdot
\dot{\vec{\Phi}}_T(-\vec{q})+\dot{\vec{\Phi}}_T(\vec{q})\cdot
\vec{\Phi}_T(-\vec{q}))
+\dot{\vec{\Phi}}_T(\vec{q})\cdot\vec{J}_T
(-\vec{q})+\vec{J}_T(\vec{q})\cdot\dot{\vec{\Phi}}_T(-\vec{q})\right]
\label{rateoperator}
\end{equation}
The expectation value of this Heisenberg operator equation in the initial
density matrix can be written in a compact notation as
\begin{eqnarray}
\langle\dot{N}_{\vec{q}}(t)\rangle
=&&-\frac{1}{2q}
\frac{\partial}{\partial t^{\prime\prime}}
\left[\langle\vec{J}^{+}_T(\vec{q},t)\cdot{\vec{\Phi}^{-}}_T
(-\vec{q},t^{\prime\prime}) 
\rangle\right]_{t=t^{\prime\prime}} \nonumber \\
 &&-\frac{\mu^2(t)}{2q}
\frac{\partial}{\partial t^{\prime\prime}}
\left[\langle\vec{\Phi}^{+}_T(\vec{q},t)\cdot{\vec{\Phi}^{-}}_T
(-\vec{q},t^{\prime\prime})+
\langle\vec{\Phi}^{+}_T(\vec{q},t^{\prime\prime})\cdot{\vec{\Phi}^{-}}_T
(-\vec{q},t)  
\rangle\right]_{t=t^{\prime\prime}} \label{ndot2}
\end{eqnarray}
where we have used fields and currents defined on different branches (forward
and backward) to avoid the potential ambiguities associated with Schwinger
terms in the time derivatives of correlation functions. The expectation value
of the emission rate can now be obtained by performing a perturbative expansion
in $\alpha$ of the expectation values above in the non-equilibrium generating
functional
\begin{equation}
\langle \dot{N}_{\vec q} \rangle
=\int{\cal D}[\vec{\Phi}^+_T]{\cal D}[\vec{\Phi}^-_T]
{\cal D}[\tilde{\pi}^+_i]{\cal D}[\tilde{\pi}^-_i]
e^{i[S_0^+-S_0^-]}e^{i[S_I^+-S_I^-]}
\dot{N}_{\vec{q}}(t)\label{expecval}
\end{equation}
with
\begin{equation}
S_I^{\pm}= -\int dt^{\prime}\sum_{\vec{p}}\left[
\frac{\mu^2(t)}{2}\vec{\Phi}^{\pm}_T(\vec{-p})
\cdot{\vec{\Phi}^{\pm}}_T(\vec{p})+ \vec{J}^{\pm}_T(\vec{-p})
\cdot{\vec{\Phi}^{\pm}}_T(\vec{p})\right]\label{inter}
\end{equation}
on the respective $\pm$ time branches. The Coulomb interaction can be
incorporated in the non-equilibrium path integral by introducing an auxiliary
field variable to replace the Coulomb interaction by a linear coupling of the
charge density to the auxiliary field (without kinetic term), and carrying out
the path integral over the canonical momenta of the pion fields. However we
omit this term since its contribution will be of higher order in $\alpha$.

Perturbation theory in $\alpha$ is now carried out in terms of the following
Green's functions for the photon field
\begin{eqnarray}
\langle\Phi^{-i}(\vec{k},t)\Phi^{+j}(-\vec{k},t^{\prime})\rangle
&=&-i{\cal{P}}^{ij}(\vec{k}){\cal{G}}^{>}_k(t,t')\nonumber\\
&=&{\cal{P}}^{ij}(\vec{k})\frac{1}{2k}\left[e^{-ik(t-t^{\prime})}(1+N_k(0))+
e^{ik(t-t^{\prime})}N_k(0)\right]\label{g1phot} \\
\langle\Phi^{+i}(\vec{k},t)\Phi^{-j}(-\vec{k},t^{\prime})\rangle
&=&-i{\cal{P}}^{ij}(\vec{k}){\cal{G}}^{<}_k(t,t^{\prime})\nonumber\\
&=&{\cal{P}}^{ij}(\vec{k})\frac{1}{2k}\left[e^{ik(t-t^{\prime})}(1+N_k(0))+
e^{-ik(t-t^{\prime})}N_k(0)\right]\\\nonumber\\
\langle\Phi^{+i}(\vec{k},t)\Phi^{+j}(-\vec{k},t^{\prime})\rangle
&=&-i{\cal{P}}^{ij}(\vec{k})\left[{\cal{G}}^{>}_k(t,t^{\prime})\Theta(t-t^{\prime})
+{\cal{G}}^{<}_k(t,t^{\prime})\Theta(t^{\prime}-t)\right] \\\nonumber\\
\langle\Phi^{-i}(\vec{k},t)\Phi^{-j}(-\vec{k},t^{\prime})\rangle
&=&-i{\cal{P}}^{ij}(\vec{k})\left[{\cal{G}}^{<}_k(t,t^{\prime})
\Theta(t-t^{\prime})+
{\cal{G}}^{>}_k(t,t^{\prime})\Theta(t^{\prime}-t)\right]\label{g2phot}\\
N_k(0)& = & \frac{1}{e^{\beta k}-1},
\end{eqnarray}
where we assume that the initial state is that of local thermodynamic
equilibrium and $N_k(0)$ is the initial photon distribution at time $t=0$. We
will consider two cases: i) the initial distribution is that of equilibrium at
an initial temperature $T_i=1/\beta$ and given by the equilibrium Bose-Einstein
distribution functions, ii) no photons in the initial state, i.e.
$N_q(0)=0$. We choose this latter initial state for comparison purposes. The
case of any arbitrary initial distribution can be treated similarly.

To zeroth order in $e$, the rate vanishes, as can be seen by taking the time
derivatives of the free-field propagators (\ref{g1phot}-\ref{g2phot}).  The
lowest order contribution to the last two terms in (\ref{ndot2}) comes from the
seagull term which is ${\cal{O}}(\alpha)$. Thus, the term proportional to
$\mu^2(t)$ will give a contribution of order $\alpha^2$, since $\mu^2(t)
\propto \alpha$, while the current vertex will yield the ${\cal{O}}(\alpha)$
leading contribution to the photoproduction rate as shown below. This means
that we need only calculate the current vertex contribution.

The first term in (\ref{ndot2}) receives contributions from the $J\cdot A_T$
term in the action. In particular, these contributions are proportional to the
hadronic (pion) current-current correlators which are ${\cal{O}}(e^2)$ and are
therefore the leading contributions to the photoproduction rate. It should be
noted, however, that in an {\em equilibrium} situation the ${\cal{O}}(\alpha)$
(lowest order) contribution from the current-current correlator vanishes due to
kinematics as it will be shown explicitly below, and the lowest order
contribution to the rate is from the seagull terms along with the Compton
scattering and pair annihilation diagrams.


\section{Photoproduction}

As mentioned above, photons and dileptons are ideal probes of the initial
stages of the phase transition as they do not undergo final state interactions
and basically freeze out with long mean free paths compared to strong
interaction length scales.  Our strategy will therefore be to develop and
implement, from first principles, a scheme for calculating the number of
photons at a given time $t$ after the quench.

The fact that the photoproduction rates are related to current-current
correlators is a well-known result. However, in this section, we will derive
such a relationship from first principles for photoproduction from a bath of
pions which is {\em far from equilibrium}.

As argued in the previous section, the lowest order contribution in $\alpha $
arises from the non-equilibrium current vertex. Therefore, to this order we
neglect both the ``seagull'' term and the Coulomb interaction. We must
emphasize that we are performing a perturbative expansion in the
electromagnetic coupling $\alpha$ , while the pion dynamics is being treated
{\em non-perturbatively}, and in fact {\em exactly} to leading order in the
$1/N$ expansion.  Thus to this order in $\alpha$,
\begin{eqnarray}
\langle\dot{N}_{\vec{q}}(t)\rangle=&&-\frac{1}{4q}
\left[\langle\dot{\vec{\Phi}}_T(\vec{q})\cdot\vec{J}_T(-\vec{q})\rangle
+\langle\vec{J}_T(\vec{q})\cdot\dot{\vec{\Phi}}_T(-\vec{q})\rangle\right]
\label{ndot1}
\end{eqnarray}
The expectation values in the above expression are expressed in terms of
correlators of operators living on the forward and backward time contours,
yielding\cite{note2}
\begin{eqnarray}
&&\langle\dot{N}_{\vec{q}}(t)\rangle\
=\frac{2ie^2}{q}
\frac{\partial}{\partial t^{\prime\prime}}
\int dt^{\prime}\int \frac{d^3k}{(2\pi)^3} k^i k^j
\left[
\langle\tilde{\pi}^+(\vec{k},t)\tilde{\pi}^+(-\vec{k},t ^{\prime})\rangle
\times \right.\\ \nonumber
&& \times
\langle\tilde{\pi}^+(-\vec{k}-\vec{q},t)
\tilde{\pi}^+(\vec{k}+\vec{q},t ^{\prime})\rangle
\langle{\Phi}^{+j}_T(-\vec{q},t^{\prime})
{\Phi}^{-i}_T(\vec{q},t^{\prime\prime})\rangle\ \\ \nonumber
&& \left.
-\langle\tilde{\pi}^+(\vec{k},t)\tilde{\pi}^-(-\vec{k},t^{\prime})\rangle
\langle\tilde{\pi}^+(-\vec{k}-\vec{q},t)
\tilde{\pi}^-(\vec{k}+\vec{q},t^{\prime})\rangle
\langle{\Phi}^{-j}_T(-\vec{q},t^{\prime})
{\Phi}^{-i}_T(\vec{q},t^{\prime\prime})\rangle \right]_{t=t^{\prime\prime}},
\label{prelimrate}
\end{eqnarray}
where we have omitted the isospin indices on the pions and used the fact that
$\langle\pi_1\pi_1\rangle=\langle\pi_2\pi_2\rangle$. The two point functions
$\langle\tilde{\pi}^+(\vec{k},t)\tilde{\pi}^-(-\vec{k},t ^{\prime})\rangle
=-iG_k^>(t,t^{\prime})$ and
$\langle\tilde{\pi}^+(\vec{k},t)\tilde{\pi}^+(-\vec{k},t ^{\prime})\rangle
=-iG_k^{++}(t,t^{\prime})$ can be read off from (\ref{gplpl}),(\ref{gfourier>})
and (\ref{gfourier<}).

Gathering all these ingredients together, we find after some lengthy but
straightforward algebra
\begin{eqnarray}
&&\langle\dot{N}_{\vec{q}}(t)\rangle\ =\frac{e^2}{4\pi^3q}\int dt^{\prime}\int
d^3k[k^2-(\vec{k}\cdot{\hat{q}})^2] 
\times
\\\nonumber\\\nonumber
&&\times\left
[G_k^>(t,t^{\prime})G_{|\vec{k}+\vec{q}|}^>(t,t^{\prime})\dot{\cal{G}}^<_q
(t^{\prime},t)-G_k^<(t,t^{\prime})G_{|\vec{k}+\vec{q}|}^<(t,t^{\prime})
\dot{\cal{G}}^>_q(t^{\prime},t)\right]\Theta(t-t^{\prime}).
\end{eqnarray}
The theta function ensures that this expressions is causal. The Green's
functions for the pions (\ref{gfourier>}), (\ref{gfourier<}) are expressed in
terms of the non-perturbative mode functions which are obtained numerically by
evolution of the equations (\ref{modeqns}).  Finally the rate can be written in
terms of the pion mode functions as (see equations
(\ref{modeqns},\ref{gfourier>},\ref{gfourier<})
\begin{eqnarray}
&&\langle\dot{N}_{\vec{q}}(t)\rangle
=\frac{e^2}{32\pi^3q}\int \frac{d^3k}{W_kW_{|\vec{k}+\vec{q}|}}k^2
\sin^2\theta\int^t_{-\infty}dt^{\prime}\label{rateqn}\\\nonumber\\\nonumber
&&\left
\{
\left[U_k(t)U_k^*(t^{\prime})U_{|\vec{k}+\vec{q}|}(t)U_{|\vec{k}+\vec{q}|}
^*(t^{\prime})
e^{-iq(t-t^{\prime})}(1+n_k)(1+n_{|\vec{k}+\vec{q}|})(1+N_q(0))
\right.\right.\\\nonumber
\\\nonumber
&&\left.\left.-U_k^*(t)U_k(t^{\prime})U_{|\vec{k}+\vec{q}|}^*(t)
U_{|\vec{k}+\vec{q}|}(t^{\prime})e^{iq(t-t^{\prime})}
n_kn_{|\vec{k}+\vec{q}|}N_q(0)\right]\right.\\\nonumber
\\\nonumber
&&\left.\hspace{0.5in}+2
\left[U_k(t)U_k^*(t^{\prime})U_{|\vec{k}+\vec{q}|}^*(t)
U_{|\vec{k}+\vec{q}|}(t^{\prime})e^{-iq(t-t^{\prime})}
(1+n_k)n_{|\vec{k}+\vec{q}|}(1+N_q(0))
\right.\right.\\\nonumber
\\\nonumber
&&\left.\left.\hspace{0.5in}-U_k^*(t)U_k(t^{\prime})U_{|\vec{k}+\vec{q}|}(t)
U_{|\vec{k}+\vec{q}|}^*(t^{\prime})
e^{iq(t-t^{\prime})}n_k(1+n_{|\vec{k}+\vec{q}|})N_q(0)\right]
\right.\\\nonumber
\\\nonumber
&&\left.\hspace{1in}+\left[U_k^*(t)U_k(t^{\prime})U_{|\vec{k}+\vec{q}|}(t)
U_{|\vec{k}+\vec{q}|}^*(t^{\prime})
e^{-iq(t-t^{\prime})}n_kn_{|\vec{k}+\vec{q}|}(1+N_q(0))
\right.\right.\\\nonumber
\\\nonumber
&&\left.\left.
\hspace{1in}-U_k(t)U_k^*(t^{\prime})U_{|\vec{k}+\vec{q}|}(t)
U_{|\vec{k}+\vec{q}|}^*(t^{\prime})
e^{iq(t-t^{\prime})}(1+n_k)(1+n_{|\vec{k}+\vec{q}|})N_q(0)
\right]+\text{c.c.}\right\}
\end{eqnarray}
There are several noteworthy aspects of this equation. First, we emphasize that to
lowest order in perturbation theory in $\alpha$, this expression is {\em exact}
in the specific sense that we have {\em not} coarse-grained over any time
scales in the problem.  All the relevant microscopic time-scales of the theory
are accounted for in this expression. This is a {\bf generalized quantum
kinetic equation}, which avoids the assumption of completed successive
scattering processes implicit in a simple Boltzmann description.  There is {\bf
no energy conservation} on short time scales ($\sim$ a fermi), but memory of 
the earlier states remains. This lack of energy conservation on short time
scales allows one to study ``transient phenomena''\cite{boylawrie,meden}.

Next, the individual terms of this expression have a very physical and simple
interpretation in terms of ``rate balancing processes''. As shown in Fig.(4.a)
the first term represents the difference in rates for two processes: the
emission and absorption of a photon and two ``off-shell'' pions in the
medium. By ``off-shell'' pions, we are referring to the pions with fully dressed
non-equilibrium propagators in the large $N$ approximation. Thus, although the
processes conserve momentum, the absence of equilibrium means that they are not
energy-conserving. The second term, depicted in Fig.(4b), represents a
``bremsstrahlung'' type process and the inverse process of ``scattering'' of a
charged pion off a photon. The reason that we identify the direct process with
a ``bremsstrahlung'' type process is because the full dressed pion propagator
has insertions of the mean-field and the self-consistent fluctuation as shown
in the figures. These processes can also be identified with a non-perturbative
resummation (large $N$) of Landau-damping type contributions, which in
equilibrium and in the usual perturbative expansion only contribute to
processes with four momenta below the light cone.  Finally the third term,
depicted in Fig. (4c) represents $\pi^+\pi^-$ annihilation into a photon, minus
the reverse process, with a photon creating a charged pion pair.

Another important point is that, as we show below, the rate vanishes
automatically if the system is in equilibrium:

In equilibrium, the pion mode functions are positive frequency plane waves
i.e. $U_k(t)=\exp({-i\omega_k t})=\exp({-i\sqrt{k^2+m^2}t})$. The time
integrals can then be easily performed by introducing an adiabatic
``switch-on'' convergence factor for the lower limit of the time integral
yielding energy conserving delta functions. In particular, we obtain
\begin{eqnarray}
&&\langle\dot{N}_{\vec{q}}(t)\rangle\
=\frac{e^2}{16\pi^2q}\int \frac{d^3k}{W_kW_{|\vec{k}+\vec{q}|}}k^2
\sin^2\theta\left\{\delta(\omega_k+\omega_{|\vec{k}+\vec{q}|}+q)\times
\right.\nonumber \\\nonumber \\
&&\left.
\times[(1+n_k)(1+n_{|\vec{k}+\vec{q}|})(1+N_q(0))-n_kn_{|\vec{k}+\vec{q}|}
N_q(0)] 
\right.
\nonumber \\\nonumber \\
&& \left.+2\delta(\omega_k-\omega_{|\vec{k}+\vec{q}|}+q)
[(1+n_k)n_{|\vec{k}+\vec{q}|}(1+N_q(0))-n_k(1+n_{|\vec{k}+\vec{q}|})
N_q(0)]
\right.
\nonumber \\\nonumber \\
&& \left. +\delta(\omega_k+\omega_{|\vec{k}+\vec{q}|}-q)
[n_kn_{|\vec{k}+\vec{q}|}(1+N_q(0))-(1+n_k)(1+n_{|\vec{k}+\vec{q}|})N_q(0)]
\right\}.
\label{equirate}
\end{eqnarray}
The delta functions can never be satisfied and so the rate simply vanishes in
equilibrium, due to the kinematics. Furthermore, upon linearizing near the
equilibrium distribution for the photons $N_q(0)=N_{eq,q}+\delta N_q$, i.e. in
the ``relaxation time approximation'', one can easily see that the expression
that multiplies $\delta N_q$ is identified with the imaginary part of the
photon self-energy to lowest order in $\alpha$ evaluated on the photon
mass-shell, which vanishes kinematically.

This is of course a well known result: the equilibrium photoproduction rate is
given by the imaginary part of the photon self-energy evaluated on the photon
mass shell.  {\em In equilibrium} the lowest order contribution to the on-shell
imaginary part of the photon self-energy arises from Compton scattering and
pair annihilation diagrams (and their reverse processes) in the medium; these
are both ${\cal{O}}(\alpha^2)$ effects.

Thus we see that {\em out of equilibrium} there are ``off-shell'' processes
that give a non-zero ${\cal{O}}(\alpha)$ contribution to photoproduction. The
equation (\ref{rateqn}) and these conclusions are some of the more important
results of this article.

Having established the equivalence with the usual result in the case in which
the state is prepared in equilibrium in the infinite past (in-state) and
evolved in time, we now consider the kinetics of photoproduction as an {\em
initial condition problem}.  In the situation under consideration, i.e.  a
sudden ``quench'' below the spinodal region in the linear sigma model, the
conclusion of the discussion above implies that if one assumes that the system
was in (local) thermodynamic equilibrium for all times before the quench $t<0$,
then the rate vanishes for $t<0$. Therefore, let us consider the following
initial value problem where the initial density matrix, prepared at time $t=0$
commutes with the respective number operators with some initial distribution of
photons and let this initial density matrix evolve in time with the total
Hamiltonian.

The calculation of the rate goes through in exactly the same manner but now the
lower limit of the retarded time integral in (\ref{rateqn}) is the initial time
$t=0$ (see appendix A for a brief discussion of kinetics as an initial value
problem and\cite{boylawrie,meden,kinetics} for more details).

\subsection{Improved Rate Equation}

In this section we present an improvement to equation (\ref{rateqn}) obtained
in the previous section, that represents a {\em resummation} of the
perturbative series in $\alpha$. The diagrammatic structure of the perturbative
series for the rate is recognized to lead to the following {\em exact}
expression for the rate

\begin{eqnarray}
&&\langle\dot{N}_{\vec{q}}(t)\rangle\
=\frac{e^2}{4\pi^3q}\int_0^t dt^{\prime}\int
d^3k[k^2-(\vec{k}\cdot{\hat{q}})^2] 
\times
\nonumber \\ \nonumber \\
&&\left
[\Sigma_{k,|\vec k +\vec q|}^>(t,t^{\prime})
\dot{\overline{\cal{G}}}^<_q
(t^{\prime},t)-\Sigma_{k,|\vec{k}+\vec{q}|}^<(t,t^{\prime})
\dot{\overline{\cal{G}}}^>_q(t^{\prime},t)\right]\Theta(t-t^{\prime}).
\label{exactrate}
\end{eqnarray}
where ${\overline{\cal{G}}}^{>,<}$ are the {\em full} photon Green's functions,
and $\Sigma^{>,<}$ are the {\em full} photon self-energies obtained to all
orders in perturbation theory in the electromagnetic coupling.  The expression
for the rate given by (\ref{rateqn}) is obtained by replacing
${\overline{\cal{G}}}^{>,<} \; ,\Sigma ^{>,<}$ by the zeroth order
contributions in terms of free field propagators. The exact expression for the
rate (\ref{exactrate}) gives the rate as a function of time and the initial
population $N_q(0)$, which combined with the Schwinger Dyson equation for the
photon propagators gives the exact rate.  The lowest order term, however,
neglects the change in the population providing only the lowest order time
evolution. A resummation must be invoked to incorporate the change of
population as a function of time.

The first point to notice is that $\dot{N}_q$ in (\ref{rateqn}) is proportional
to the fine structure constant $\alpha$. The presence of this weak coupling is
a signal that there is a separation of time scales in the problem. To be
specific, the time scales we are referring to are: (i) $(\dot{N}_q/N_q)^{-1}$
which is the time scale that governs photoproduction; and (ii) the time scale
associated with the non-equilibrium processes $\sim$ 1 fm/c which are the time
scales for evolution of the pion mode functions. The presence of an $e^2$ in
the expression for $\dot{N}_q$, guarantees that these two time scales are
widely separated. This means that the photon distribution will show only small
deviations from equilibrium on short time scales. Obviously these expressions
cannot be extended to long time scales since they neglect the changes in the
initial distribution. 

Now consider implementing the following procedure. Let us integrate
(\ref{rateqn}) from some initial time $t_0$ to a time $t_0+\Delta t$, (where
$\Delta t<$ $1$fm/c, i.e. the microscopic time scale), hence obtaining
$N_q(t_0+\Delta t)$. Assuming that the change in the photon distribution
function is small during this interval (an assumption warranted for weak
coupling), we can then update the value of $N_q$ that enters into the photon
Green's functions on the RHS of (\ref{rateqn}) to $N_q(t_0+\Delta t)$. In terms
of the density matrix, this procedure has the interpretation of starting off
with a density matrix that is diagonal in the number basis. As the system
evolves, correlations of the form $\langle b_q(t)b_{-q}(t) \rangle$ and
$\langle b_q^{\dagger}(t)b_{-q}^{\dagger}(t) \rangle$ are generated, which give
rise to off-diagonal terms in the density matrix \cite{boylawrie}. The updating
procedure outlined above neglects these correlations by collapsing the density
matrix to a diagonal one at the end of each infinitesimal time step.

This procedure is iterated for all times, resulting in the replacement $N_q(0)
\rightarrow N_q(t^{\prime})$. This approximation is similar to that employed in
quantum kinetics, known as the generalised Kadanoff-Baym
approximation\cite{meden}. It can be seen to sum a Dyson-like series for the
rate\cite{kinetics} by writing the formal solution iteratively.

A similar resummation scheme is implied by the semiclassical Boltzmann
equation, in which if the occupation numbers are treated in lowest order, the
change is linear in time. Replacing the occupation numbers by the time
dependent ones in the Boltzmann equation leads to a resummation and
exponentiation of the time series\cite{boylawrie}. However, as discussed
in\cite{boylawrie} the Boltzmann equation assumes completed collisions that
result in a coarse graining in time and neglects all of the transient effects
and dynamics on short time scales.  See appendix A and
\cite{boylawrie,meden,kinetics} for a more detailed discussion of the
approximations involved.

We note that the pion occupation numbers $n_k$ are not updated since the change
in the occupation numbers for the pions is accounted for by the evolution of
the mode functions through the Bogoliubov transformation that determines the
time evolution of the particle number\cite{boyan}.

Since we are using a ``box'' normalization for the particle states, we now pass
to the ``continuum'' normalization (in a volume $\Omega$) by the replacement
\begin{equation}
\langle N_q(t) \rangle \rightarrow 
(2\pi)^3 \frac{d(\langle N(t) \rangle/\Omega)}{d^3q} \label{cont1}
\end{equation}
\begin{equation}
 \langle \dot{N}_q(t) \rangle \rightarrow 
(2\pi)^3 \frac{d(\langle \dot{N}(t) \rangle/\Omega)}{d^3q}
\equiv (2\pi)^3 \frac{dR(t)}{d^3q}\label{contrate}
\end{equation}
and obtain the final form of the invariant rate of photoproduction {\em per
polarization}:
\begin{eqnarray}
&&(2\pi)^3 |q|\frac{dR(t)}{d^3q}
=\frac{\alpha}{8\pi^2}\int \frac{d^3k}{W_kW_{|\vec{k}+\vec{q}|}}k^2
\sin^2\theta\int^t_{0}dt^{\prime}\nonumber \\
\nonumber \\
&&\left
\{
\left[U_k(t)U_k^*(t^{\prime})U_{|\vec{k}+\vec{q}|}(t)U_{|\vec{k}+\vec{q}|}
^*(t^{\prime})
e^{-iq(t-t^{\prime})}(1+n_k)(1+n_{|\vec{k}+\vec{q}|})(1+N_q(t^{\prime}))
\right.\right.\nonumber \\\nonumber \\
&&\left.\left.-U_k^*(t)U_k(t^{\prime})U_{|\vec{k}+\vec{q}|}^*(t)
U_{|\vec{k}+\vec{q}|}(t^{\prime})e^{iq(t-t^{\prime})}
n_kn_{|\vec{k}+\vec{q}|}N_q(t^{\prime})\right]\right.\nonumber \\\nonumber \\
&&\left.\hspace{0.5in}+2
\left[U_k(t)U_k^*(t^{\prime})U_{|\vec{k}+\vec{q}|}^*(t)
U_{|\vec{k}+\vec{q}|}(t^{\prime})e^{-iq(t-t^{\prime})}
(1+n_k)n_{|\vec{k}+\vec{q}|}(1+N_q(t^{\prime}))
\right.\right.\nonumber \\\nonumber \\
&&\left.\left.\hspace{0.5in}-U_k^*(t)U_k(t^{\prime})U_{|\vec{k}+\vec{q}|}(t)
U_{|\vec{k}+\vec{q}|}^*(t^{\prime})
e^{iq(t-t^{\prime})}n_k(1+n_{|\vec{k}+\vec{q}|})N_q(t^{\prime})\right]
\right.\nonumber \\\nonumber \\
&&\left.\hspace{1in}+\left[U_k^*(t)U_k(t^{\prime})U_{|\vec{k}+\vec{q}|}(t)
U_{|\vec{k}+\vec{q}|}^*(t^{\prime})
e^{-iq(t-t^{\prime})}n_kn_{|\vec{k}+\vec{q}|}(1+N_q(t^{\prime}))
\right.\right.\nonumber \\\nonumber \\
&&\left.\left.
\hspace{1in}-U_k(t)U_k^*(t^{\prime})U_{|\vec{k}+\vec{q}|}(t)
U_{|\vec{k}+\vec{q}|}^*(t^{\prime})
e^{iq(t-t^{\prime})}(1+n_k)(1+n_{|\vec{k}+\vec{q}|})N_q(t^{\prime})
\right]+\text{c.c.}\right\},\label{finalrate}
\end{eqnarray}
where $\theta$ is the angle between $\vec k$ and $\vec q$.  This novel
expression for the rate is the basic important result of this article. The
total photon yield at a time $t$ is given by twice (to account for both
polarizations) the time integral of (\ref{finalrate}).

\section{Numerical study and results:}

Having established the proper rate equation that accounts for strong
off-equilibrium effects we now proceed to a numerical evaluation of the rate
and the total number of photons produced during the time of the transition.

The numerical study consists of two stages: in the first stage the evolution of
the order parameter and the mode functions is solved by integrating the
equations (\ref{zeromodeeqn},\ref{modeqns}, \ref{inicon},\ref{fluct}) with
(\ref{massoft1}) and $N=3$ corresponding to 3 pion degrees of freedom. We have
chosen to represent a ``quench'' from an initial temperature $T_i=1.1 T_c=220
$ MeV to zero temperature. This choice has no particular physical significance
but serves as an illustration of a quench scenario not too far above the
critical temperature.  We have also performed calculations with the initial
temperature approching the critical temperature from above.

This stage of the numerical evaluation produces the evolution of the order
parameter and provides the mode functions $U_k(t)$ for all the values of $k$
considered. Consistent with the linear sigma model being an effective theory
below 1 GeV, we have kept all $k$ wave vectors up to this value.  As shown in
figure(1) the order parameter reaches the saturation value $\approx f_{\pi}$ on
time scales $ \approx 3$ fm/c, this time scale signals the last stages of the
phase transition. Thus we kept all the mode functions up to this time.

The second stage of the calculation uses the mode functions as input in the
numerical evaluation of the rate expression (\ref{finalrate}).  The input for the
distribution functions is taken to be an equilibrium distribution of photons
and pions at the initial temperature which is then varied from $T_i=1.1 T_c$,
down to $T_c$.  As we perform the calculation of the rate (\ref{finalrate}), we
also simultaneously integrate the expression to obtain the total number of
photons per unit volume at a given time (\ref{cont1}) (multiplied by $|q|$).

\vspace{2mm}

{\bf Results:} 

The results of the numerical evaluation of the rate are clearly displayed in
figures (5-8).  Figure(5.a) shows the total photon yield (i.e. for both
polarizations) per unit volume $(2\pi)^3 |q|\frac{d(\langle N(t)
\rangle/\Omega)}{d^3q}$ (in units of $\mbox{fm}^{-1}$) at time $t=$3\ fm/c vs
$|q|$ (in units of 200 MeV) . We clearly see that the distributions of photons
produced during the time of the phase transition is out of equilibrium.  The
long wavelength photons, with energies $\leq 100$ MeV can be described by a
thermal distribution at a temperature $T_{lw} \approx 1.17 \ T_c$.  The
distribution for short wavelength photons is basically not modified from the
initial distribution and merges with the initial thermal distribution at
$T_{sw}=1.1\ T_c$. The distribution function smoothly interpolates between a
thermal distribution at a temperature which is about $ 6 \%$ above the initial
temperature at long wavelengths, and another thermal distribution at the
initial temperature at short wavelengths, with a smooth crossover at energies
$\approx 100$ MeV.  

To better quantify the production of photons, we show in figure (5.b) the
difference between the distribution at $t=3$ fm/c and the initial distribution,
thus subtracting the thermal background.  The enhancement at low momenta is
clearly displayed in this figure where it can be seen that the effect is
definitely more marked for $k < 100$ MeV.  Figure (6) shows the invariant rate
$(2\pi)^3 |q| dR(t)/d^3q$ (in units of $\mbox{fm}^{-2}$) at time $t=3$ fm/c vs
$|q|$ (in units of 200 MeV). Again, clearly the rate is enhanced at very low
momenta, in the same range as shown in Figure (5.b) thus explaining the
enhancement in the total photon yield in this energy range.  The reason for
this dramatic effect at long-wavelengths is physically clear to understand.
The pion mode functions that enter in the expression of the rate
(\ref{finalrate}) grow exponentially because of the spinodal instabilities for
$k \leq 100 $ MeV as shown in figure (2.a).  Large transferred photon momentum
$\vec q$ takes the mode functions outside the band of spinodally unstable modes
with a much smaller contribution to the rate and total photon yield. This
phenomenon is quite independent of the initial distribution function for the
photons and solely a feature of the instabilities associated with the chiral
phase transition.

To see that this result is independent of the initial distribution of photons
we show in figure (7) $(2\pi)^3 |q|\frac{d(\langle N(t) \rangle/\Omega)}{d^3q}$
at time 3 fm/c for the case of initial ``vacuum'' conditions on the photon
occupation $N_q(0)=0$. This figure is very similar to the subtracted result
displayed in figure (5.b), thus explicitly showing that the enhancement is
solely a consequence of the pion instabilities and independent of the initial
photon distribution.  Figure (8) shows the invariant rate for this initial
condition.

A further enhancement of the photoproduction rate and total photon yield is
obtained with quenches closer to the critical temperature, with a dramatic
enhancement of about $15-20\%$ at low momentum when the quench is at the
critical temperature. This is an understandable result because at the critical
temperature pions are effectively massless with large contribution from their
equilibrium distribution functions at long wavelengths.  However, we find that
a scenario based on a critical quench is physically quite unlikely.

At this stage it is very illuminating to compare our results with those
obtained for photoproduction from a hadronic gas by Kapusta, Lichard and
Seibert\cite{lichard} and more recently by Steele, Yamagishi and
Zahed\cite{zahed}.  In reference\cite{lichard} the result for the invariant
rate of photoproduction from $q\bar{q}\rightarrow \gamma g \; ; \; q(\bar{q})g
\rightarrow q(\bar{q})g \gamma$ is given by
\begin{equation}
(2\pi)^3 E \frac{dR}{d^3q} = \frac{20 \pi}{9} \alpha \alpha_s T^2
e^{-E/T} \ln\left(\frac{2.9 E}{4\pi \alpha_s T}+1\right)
\end{equation}
which for $E \leq 100 \mbox{ Mev }$ and $T \approx 200 \mbox{ Mev }$ is at
least an order of magnitude smaller than the results shown in figures (6,8).
In figure 5 of reference\cite{zahed} the invariant rate for photoproduction at
a temperature 150 MeV is shown.  Although their initial temperature is smaller
than the value used by us, we can see that after normalization of scales and
units, the rate that we find for momenta $q < 100$ MeV (shown in figures (6,8))
is several orders of magnitude {\em larger} than that displayed in that
reference. Thus the main conclusion of this comparison: the long wavelength
instabilities associated with the fast phase transition below the critical
temperature are responsible for a dramatic enhancement of the photoproduction
rate and yield for low momenta.

\section{Contribution from the anomalous decay $\pi^0 \rightarrow 2\gamma$:}

One of the most important sources of low energy photons in a typical heavy ion
collision is the decay of the neutral pion into two photons, with a branching
ratio of almost $99\%$. This process typically produces photons with energies
$\geq 70$ MeV.

Thus it is important that we quantify the contribution of $\pi^0$ decay to the
non-equilibrium production of photons during the phase transition.  It is
straightforward to perform the calculation leading to the rate of
photoproduction, now including the anomalous vertex given by
(\ref{anomalousvertex}). The lowest order contribution is of
${\cal{O}}(\alpha^2)$, because each anomalous vertex is of ${\cal{O}}(\alpha)$,
and depicted in figure (9). Notice that the intermediate state has one pion
propagator, therefore {\em two} mode functions $U_k(t)$. This contribution must
be compared to the one-loop contribution depicted in figure (3), which is of
${\cal{O}}(\alpha)$ and the intermediate state has two pion propagators with
{\em four} mode functions.  The important point to notice is that the
exponentially growing (spinodally unstable) modes will give a much larger
contribution to the diagram with the pion loop than to the diagram for $\pi_0$
decay. Furthermore there is a factor of $\alpha$ difference between the two
giving an even smaller weight to the neutral pion decay diagram in favor of the
one-loop pion diagram.

Thus we conclude that the non-equilibrium process described in the
previous section gives a far larger contribution to the photoproduction
rate {\em during} the time of the phase transition than the contribution
from neutral pion decay. Both because of the fast non-equilibrium
growth of spinodally unstable modes and because of the powers of $\alpha$.

We are then led unequivocally to the conclusion that the non-equilibrium
photoproduction process through spinodal instabilities is far more
efficient than $\pi^0 \rightarrow 2 \gamma$ for photons with energy
below 80 MeV.

\section{Discussion, conclusions and outlook:}

In this article we have focused on the description of the process of
photoproduction during the non-equilibrium stages of the chiral phase
transition and formation of chiral condensates. The premise of this work is
that if the chiral phase transition occurs far from equilibrium, resulting in
the possible formation and relaxation of disoriented chiral condensates, the
long wavelength pion instabilities will lead to an enhancement of
photoproduction at low energies.

We have developed a novel quantum kinetic approach to photoproduction that
accounts for the non-equilibrium dynamics in short time scales. This approach
incorporates consistently the dynamics of long-wavelength pion fluctuations
that undergo spinodal instabilities during the phase transition. These
instabilities lead to an enhanced photoproduction rate and photoproduction yield
at low energies $|q| \leq 80 $ MeV. Comparing our results with recent results
of photoproduction in the hadronic gas\cite{lichard,zahed}, we are led to conclude that
{\em if} the chiral phase transition occurs far off equilibrium there will be a
dramatic enhancement in photoproduction in the energy range below 80-100
MeV. This result is independent of the initial photon distribution, and its
origin resides solely on the strong instabilities of the pion fluctuations, and
thus on the dynamics of the chiral phase transition.

Although we have focused on the ``quench'' scenario, our quantum kinetic
approach is certainly more general and can be generalized to contemplate the
case in which the transition occurs by bubble nucleation as proposed
recently\cite{kapvis}. Our approach did not consider either longitudinal or
spherical expansion, since we were interested in understanding if new phenomena
could emerge. The next step is to extend these methods to include
expansion\cite{cooper} as well as inhomogeneous configurations. In particular,
one can now study the fluctuations around semiclassical configurations, such as
the Blaizot-Krzywicki DCC and perform a calculation along the lines detailed in
this article that would include both the classical currents and the quantum
fluctuations, thus extending the results of\cite{wang} to include the full
quantum and thermal evolution.

Another avenue to pursue is a detailed computation of dilepton rates, by
extending the kinetic approach developed here, also incorporating the effects
of hydrodynamic expansion. The ultimate goal of such a program is to offer
detailed probes of the dynamics of the chiral phase transition either by
revealing the formation and relaxation of Disoriented Chiral Condensates or
some other type of non-equilibrium phenomena associated with the chiral phase
transition. The energy range in which the enhancement occurs is a difficult one
for the present detectors at AGS and SPS. However both the full acceptance
detector (FELIX) proposed at CERN-LHC and the PHENIX detector scheduled to
operate at RHIC towards the end of the millenium may provide the necessary
resolution at low energies to provide a window to probe these phenomena.

\acknowledgements D. B. would like to thank R. Pisarski, E. Mottola,
K. Rajagopal, J. Randrup, J. Bjorken, C. Taylor, F. Cooper, Y. Kluger,
J.P. Blaizot, X.-N. Wang and W. Cleland for illuminating discussions and
comments. He gratefully acknowledges support from N.S.F. through grant awards:
PHY-9302534 and INT-9216755, the Pittsburgh Supercomputer Center for grant
award No: PHY950011P, and to LPTHE for warm hospitality. D. B. and H. J. de
Vega thank the organizers of the Workshop on Disoriented Chiral Condensates at
ECT in Trento for a stimulating environment. R.H. and S.P.K were supported in
part by DOE Grant $\#$ DE-FG02-91-ER40682.

\appendix

\section{Formal kinetic equation}
Our treatment of photoproduction far off equilibrium is based on a description
of kinetics as an initial condition problem. Such an approach, although
non-standard in field theory at finite temperatures, has become a standard in
the studies of fast processes in semiconductor physics\cite{meden}. Far off
equilibrium situations result in solid state physics when semiconductors (or
other materials) are studied with fast pulsed lasers, which probe the dynamics
of these systems on femtosecond scales.  It has been recognized in the last few
years that a Boltzmann approach that coarse-grains over the microscopic time
scales misses most of the important strongly out of equilibrium effects
associated with virtual transitions so that the full quantum kinetic equations
must be studied. These equations are typically non-Markovian (with memory
kernels) and non-linear, to allow for non-linear relaxation, thus going well
beyond the relaxation time approximation usually invoked in near-equilibrium
situations.

A formal kinetic equation for an initial condition problem is obtained as
follows. Consider that at some initial time $t=t_o$ the system is described by
a given density matrix $\hat{\rho}(t_o)$ that commutes with the number operator
$N$, but evolves in time with a Hamiltonian that can depend explicitly on time
(as is the case under consideration in this article). The time dependent
number operator in the Heisenberg picture obeys
\begin{equation}
\dot{N}(t)= i \left[H(t),N(t)\right] \label{heis}
\end{equation}
with solution
\begin{equation}
N(t)= N(t_o)+ i\int_{t_o}^t \left[H(t'),N(t')\right]dt' \label{itera}
\end{equation}
This equation leads to an iterative series for $N(t)$ in terms of
higher order commutators and $N(t_o)$. Iterating this equation once,
and using the assumption that the initial density matrix commutes with
the number operator at the initial time, one obtains the
following {\em exact} expression for 
 the expectation value of the rate in the
initial density matrix,  
\begin{equation}
\langle \dot{N}(t) \rangle = - \int_{t_0}^t \langle
\left[H(t),\left[H(t'),N(t')\right]\right]\rangle dt' \label{ratexact}
\end{equation} 

The lowest order contribution to the rate results by replacing $N(t')$ by
$N(t_o)$. The expression (\ref{ratexact}) reveals two important features of the
exact expectation value of the rate: i) the rate vanishes at $t=t_o$ as a
result of the assumption that the initial density matrix commutes with the
number operator at the initial time, ii) the number operator enters formally in
the expression above with the time argument that is being integrated. This
expression reveals a non-Markovian structure in the rate equation in terms of
the memory kernels resulting from the nested commutators, and the argument of
the operator number is always integrated in the retarded time integrals. This
observation leads to the generalized Kadanoff Baym approximation\cite{meden},
in which the calculation of the rate is performed to first order in
perturbation theory, corresponding to replacing $N(t')=N(t_o)$ in
(\ref{ratexact}), recognizing the proper kernel from this expression (the
double nested commutator) and finally replacing the expectation values of the
number operator at the initial time by the expectation values at the integrated
times $t'$. By expanding the expectation values in terms of number eigenstates
and inserting the identity in terms of these states one finds that this
approximation neglects higher order correlations, in particular off-diagonal
matrix elements of the time dependent number operator in the basis of the
number operator at the initial time as discussed in section IV. By iterating
the resulting expression one finds that this approximation results in a
resummation of the terms akin to the Dyson series. In fact in equilibrium it
can be shown\cite{kinetics} that such a resummation is {\em precisely} a
Dyson-type approximation.  Such a resummation is also implied in the usual
Boltzmann equation, however, in order to obtain the Boltzmann equation more
drastic approximations must be made. They correspond to setting $N(t^{\prime})
\approx N(t)$, taking the contribution from the $N(t)$ outside the time
integral and performing the time integral up to $t \rightarrow \infty$
enforcing energy conservation {\em a l\'a} Fermi's Golden Rule. Such a coarse
graining approximation is refered to as ``the completed collision
approximation'' in the language of quantum kinetics. This approximation
completely neglects dynamics on short time scales and transient effects. It is
known to fail when strongly out of equilibrium processes, ocurring on short
time scales are important\cite{meden}.

Keeping the memory kernels results in a partial resummation of the exact rate
equation akin to the resummation of particular diagrams in the Dyson series for
self energies and allows to study ``transient'' phenomena associated with the
initial stages of relaxation of the initial state \cite{kinetics} which are
completely missed in a Boltzmann description, which leads always to exponential
relaxation through energy conserving processes.

{\bf Two numerical strategies:} 

The typical rate equations of quantum kinetics obtained in this article
(\ref{finalrate}) are non-Markovian and therefore non-local in time. There are
two ways to deal numerically with these : i) carry out the retarded integral
directly performing all the momentum integrations or ii) recognize that the
rate can be written in the generalized form of a sum of factorized kernels
\begin{equation}
\dot{N} = \int^t_0 dt' \sum_k \gamma_k(t) \kappa_k(t') =
\gamma_k(t) H_k(t)
\end{equation}
\noindent with the supplementary variables $H_k(t)$ that obey the
differential equations
\begin{equation}
\frac{dH_k(t)}{dt} = \kappa_k(t) \; ; \; H_k(0)=0
\end{equation}
Thus the non-local Markovian kernel is traded for a set of local
first order differential equations.   

\section{A pedagogical exercise: the forced harmonic oscillator}

Since the treatment of kinetics as an initial value problem is not part of the
standard lore, we present in this appendix a pedagogical exercise with the
purpose of providing the reader a simpler setting within which to understand
these concepts and that captures the essence of the scheme that is used in the
field theory calculation. The system that we consider is a simple harmonic
oscillator coupled to a {\em classical} source. This is already a significant
departure from the actual problem, where the electromagnetic field couples to
the charged pion current which is a quantum-mechanical object, but will
allow us to make contact with standard resuls. Thus we
proceed with our example which is introduced purely for illustrative
purposes. The Hamiltonian for our system is
\begin{equation}
H=\frac{\hat{p}^2}{2}+ \frac{\omega^2\hat{q}^2}{2}+j(t)\hat{q}
\end{equation}
where $\hat{q}$ and $\hat{p}$ are the canonical coordinate and momenta
respectively, while $\omega$ is the oscillator frequency. The number operator
for the oscillator quanta is given by
\begin{equation}
\hat{N}=\frac{1}{\omega}\left[\frac{\hat{p}^2}{2}+ \frac{\omega^2\hat{q}^2}
{2}-\frac{1}{2}\right].
\end{equation}
Therefore,
\begin{equation}
\frac{d\hat{N}}{dt}=\frac{1}{2\omega}\left[\hat{p}\dot{\hat{p}}+\dot{\hat{p}}
\hat{p}+\omega^2\hat{q}\dot{\hat{q}}+\omega^2\dot{\hat{q}}\hat{q}\right].
\end{equation}
Using the Heisenberg equations of motion, namely
\begin{eqnarray}
\dot{\hat{q}}&=&p,\\
\dot{\hat{p}}&=&-\omega^2\hat{q}-j(t)\label{hameq}
\end{eqnarray}
we obtain,
\begin{eqnarray}
\dot{\hat{N}}&=&-\frac{j(t)}{\omega}\hat{p}\\\nonumber
&=&-\frac{j(t)}{\omega}\dot{\hat{q}}.
\end{eqnarray}
>From the equations of motion (\ref{hameq}) we see that the Heisenberg operator
$\hat{q}$ satisfies the equation
\begin{equation}
\ddot{\hat{q}}+\omega^2\hat{q}=-j(t).
\end{equation}
The solutions to the above equation can be written in terms of a homogeneous
solution to the operator equation and a c-number piece
\begin{equation}
\hat{q}(t)=\hat{q}_0(t)+\int G_{ret}(t-t^{\prime})j(t^{\prime})dt^{\prime}
\end{equation}
where $G_{ret}$ is the retarded Green's function
\begin{equation}
G_{ret}(t-t^{\prime})=\frac{1}{\omega}\sin[\omega(t-t^{\prime})]
\Theta(t-t^{\prime}).
\end{equation}
We now compute operator expectation values in an initial equilibrium state
specified at $t=t_0$ which we chose to be the ground state of the unperturbed
harmonic oscillator in which $\langle \hat{p}\rangle
=\langle\dot{\hat{q}}\rangle =0$.The current is switched on at time $t=t_0$.
Considering this initial state, the expectation value of the number operator in
the time evolved state in the presence of the current is given by
\begin{equation}
\langle\dot{\hat{N}}\rangle=-\frac{j(t)}{\omega}
\langle\dot{\hat{q}}\rangle 
=\frac{j(t)}{\omega}\int^t_{t_0}
dt^{\prime}\cos[\omega(t-t^{\prime})]j(t^\prime) 
\end{equation}
leading to the final result for the number of quanta produced up to time $t$
\begin{equation}
\langle\hat{N}(t)\rangle=
\frac{1}{\omega}\int_{t_0}^tdt_1\int_{t_0}^{t_1} dt_2
j(t_1) \cos[\omega(t_1-t_2) ]j(t_2)  \label{totnum}
\end{equation} 
We can evaluate the same expectation value in the closed time path formalism
outlined in the previous section i.e.
\begin{equation}
\langle\dot{\hat{N}}\rangle=-\frac{j(t)}{\omega}\frac{d}{dt}
\int{\cal D}[q^+]{\cal D}[q^-]q^+(t)e^{\int dt^{\prime}[L^+-L^-]}
\end{equation}
where the lagrangian is 
\begin{equation}
L=\frac{1}{2}\dot{q}^2-\frac{1}{2}\omega^2{q}^2-j(t)q.
\end{equation} 
Therefore,
\begin{equation}
\langle\dot{\hat{N}}\rangle=-\frac{j(t)}{\omega}\frac{d}{dt}
\langle q^+(t)e^{-i\int dt^{\prime}j(t^{\prime})(q^+(t^{\prime})
-q^-(t^{\prime}))}\rangle_0.
\end{equation}
Expanding out the exponential and imposing the tadpole condition i.e. $\langle
q^+ \rangle_0=0$, we see that the only non-zero contribution is from the
first-order term, which yields:
\begin{equation}
\langle\dot{\hat{N}}\rangle=i\frac{j(t)}{\omega}\frac{d}{dt}
\int^{\infty}_{t_0} dt^{\prime}j(t^{\prime})(\langle
q^+(t)q^+(t^{\prime})\rangle_0 
-\langle q^+(t)q^-(t^{\prime}\rangle_0).
\end{equation}
The two point functions can be read off from (\ref{gplpl}), (\ref{gplmin}) and
(\ref{gfourier>}), giving
\begin{eqnarray} 
\langle\dot{\hat{N}}\rangle&=&i\frac{j(t)}{2\omega^2}\frac{d}{dt}
\int^t_{t_0} dt^{\prime}j(t^{\prime})(e^{-i\omega(t-t^{\prime})}
-e^{i\omega(t-t^{\prime})})\nonumber\\
&=&\frac{j(t)}{\omega}
\int^t_{t_0} dt^{\prime}j(t^{\prime})\cos[\omega(t-t^{\prime})].
\end{eqnarray}
Thus first order perturbation theory gives us the {\em exact} result. This is
of course due to the fact that the current is a c-number object.

At this point we can make contact with the usual in-out approach based on
S-matrix theory to calculate rates. Consider the expression (\ref{totnum}) in
the limit $t_0 \rightarrow -\infty$, the total number of quanta created at $t
\rightarrow \infty$ is given by
\begin{equation}
\langle \hat{N}(\infty)\rangle = \frac{|\tilde{j}(\omega)|^2}{2\omega}
\end{equation}
with $\tilde{j}(\omega)$ being the Fourier transform of the current evaluated
at the oscillator frequency $\omega$. This is the standard result for the case
of a non-interacting theory in the presence of a c-number current and makes the
connection with the treatment of Huang and Wang\cite{wang} for the sigma model.

\newpage
\begin{figure}
\epsfig{file=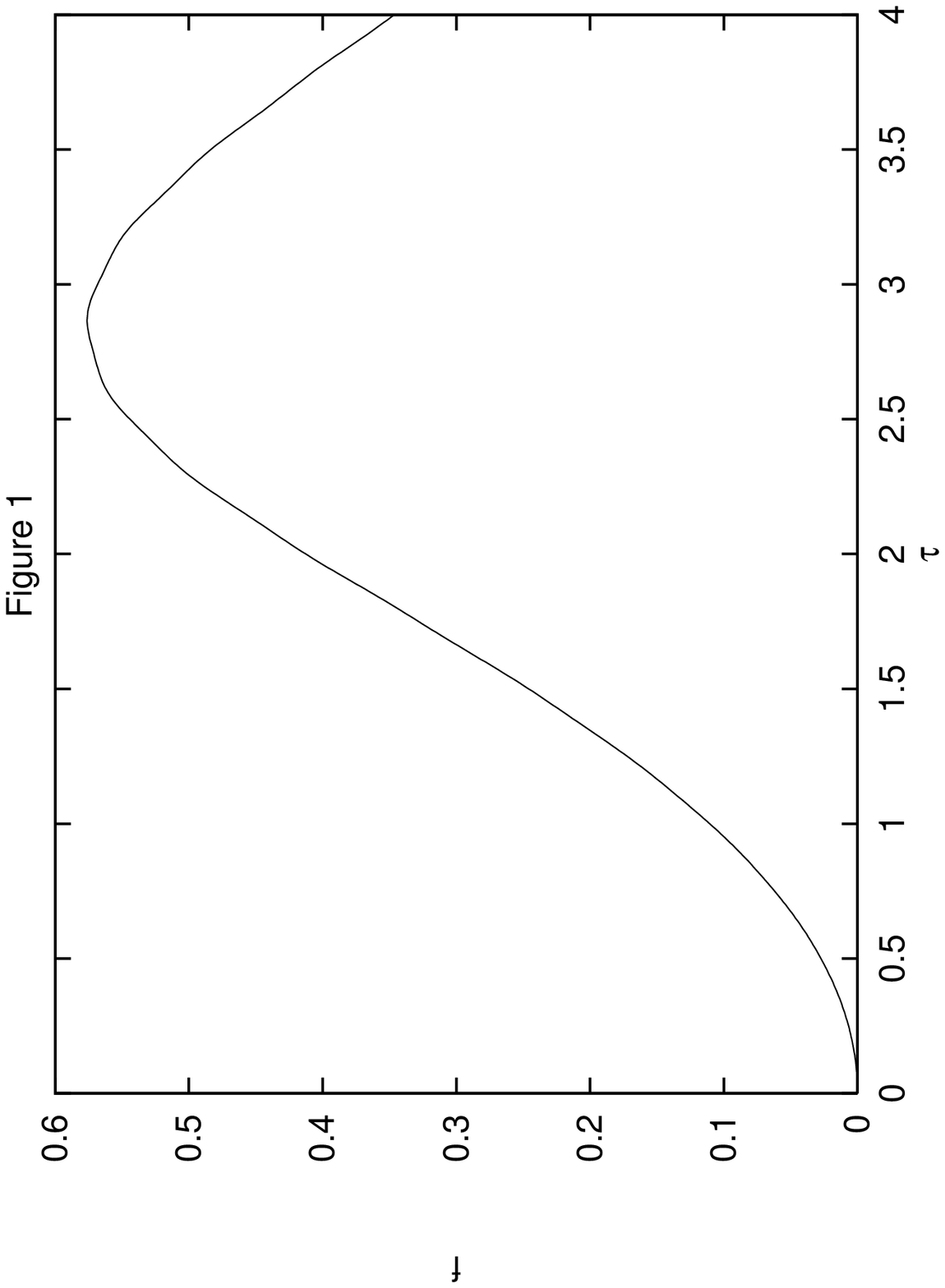}
\caption{Evolution of the order parameter $f(\tau)= \phi(t)/M_F$ vs 
$\tau = cM_F t$ ($M_F=200 \mbox{Mev}= 1/ \mbox{fm}$) for the initial 
conditions $\phi(0)=\dot{\phi}(0)=0$.}
\epsfig{file=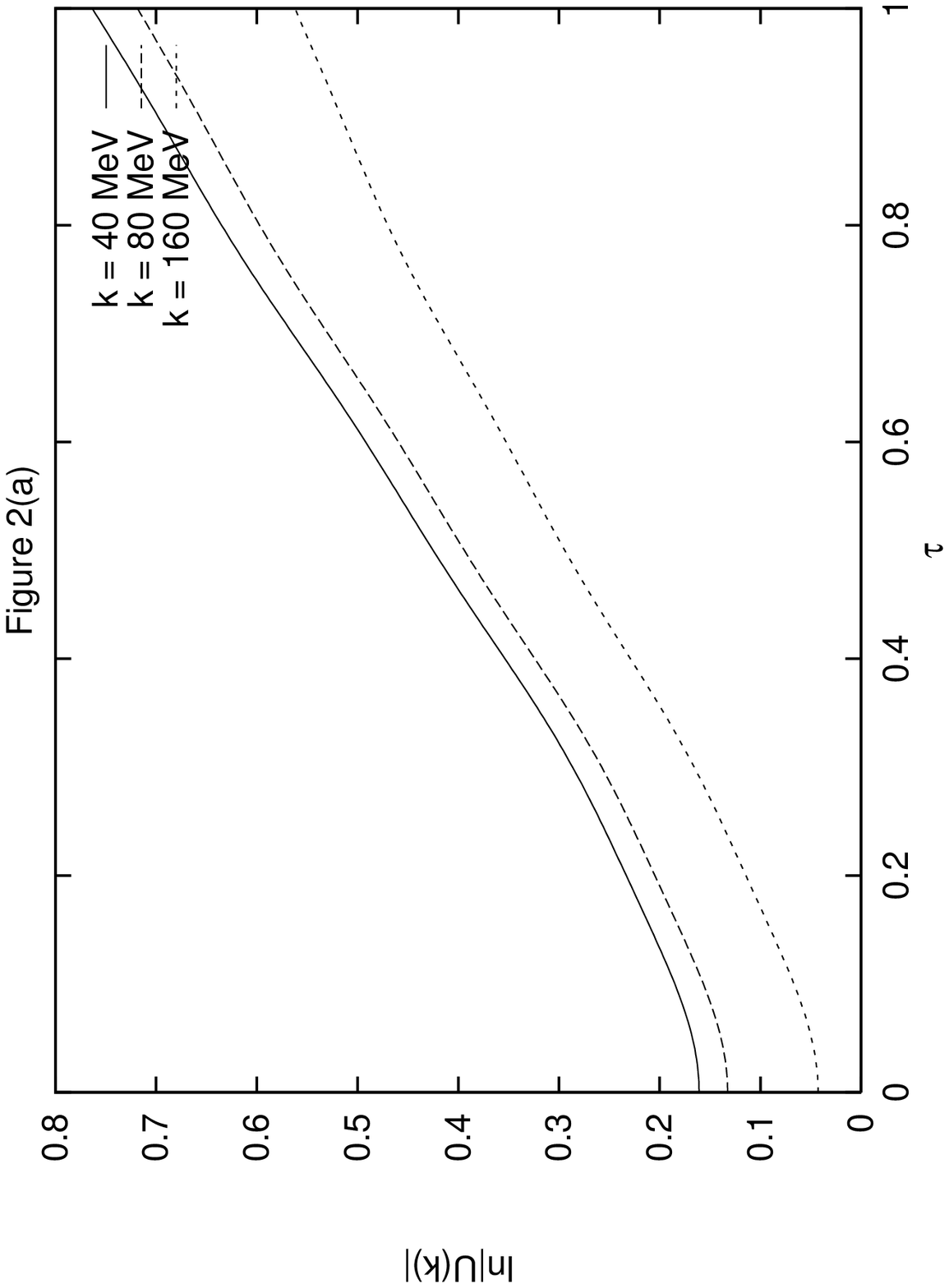}
\caption{(a) The logarithm of the modulus of the mode functions,
$\ln(|U_k(t)|)$ vs $\tau$ (time in units of fm/c) for several values of $k$.}
\addtocounter{figure}{-1}
\epsfig{file=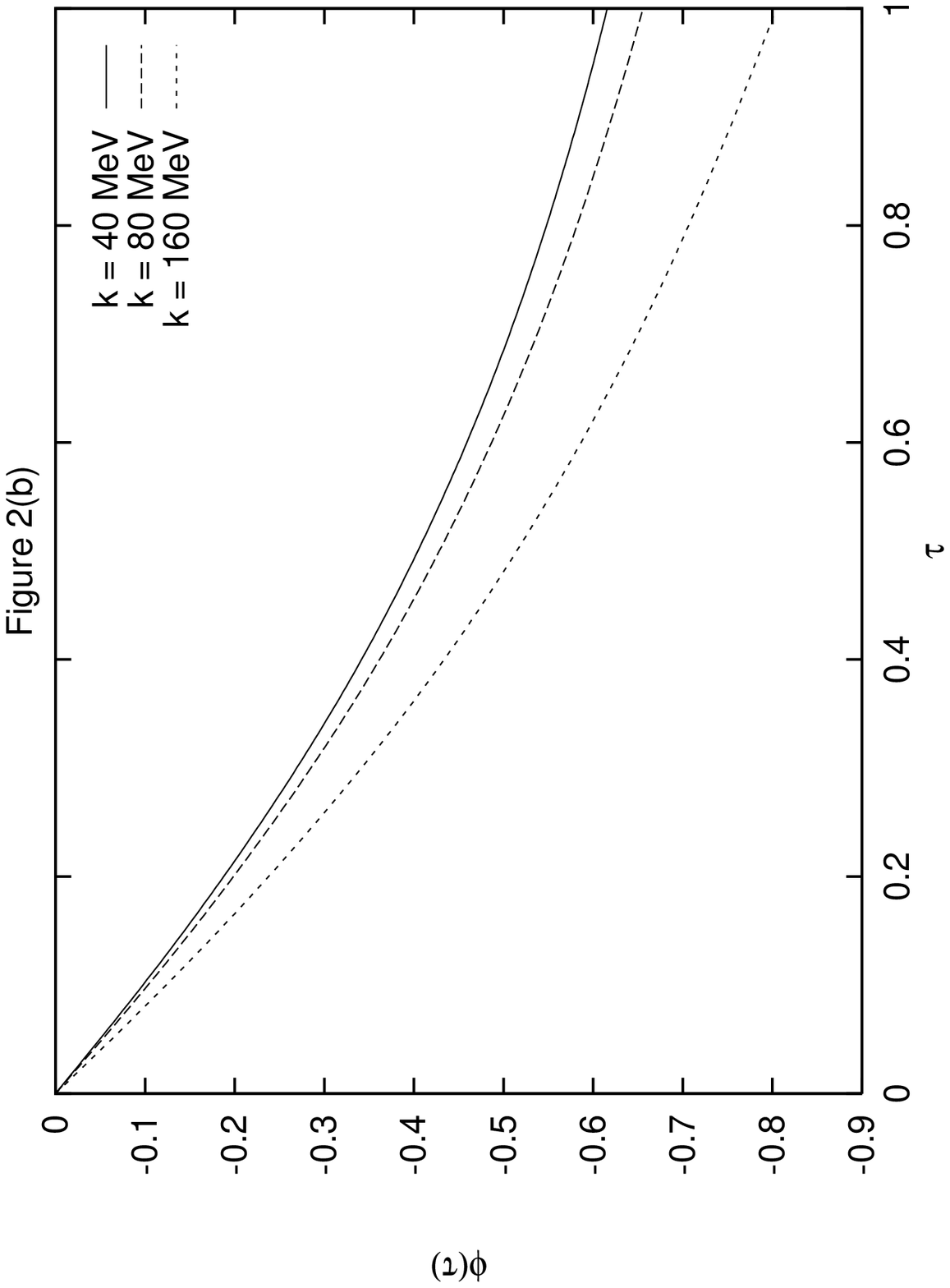}
\caption{(b) The phases of the mode functions $\varphi_k(t)$ vs $\tau$ 
for the same values of $k$ as in Figure 2.a.}
\epsfig{file=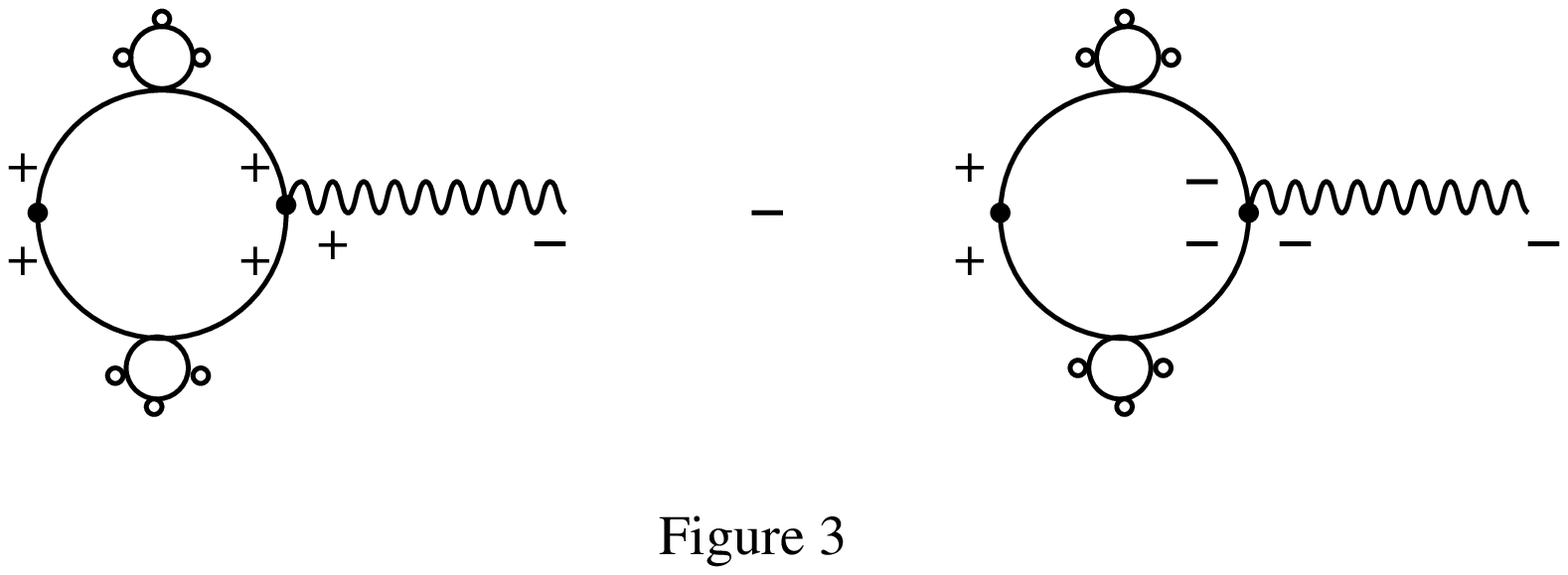}
\caption{Diagrams contributing to ${\cal{O}}(\alpha)$ to the non-equilibrium
photoproduction rate.}
\newpage
\epsfig{file=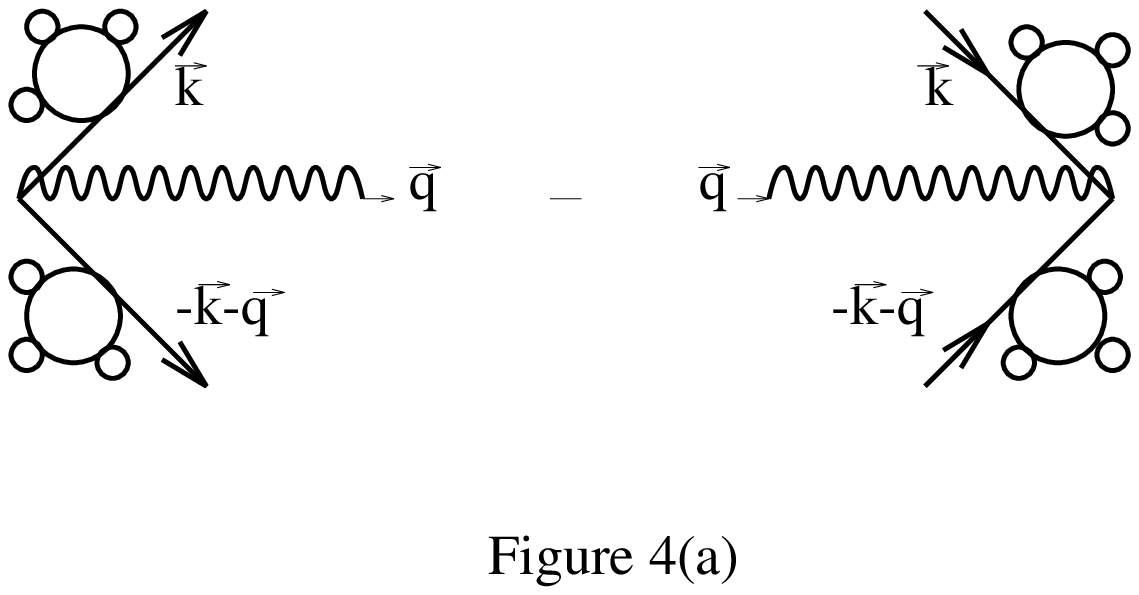}
\epsfig{file=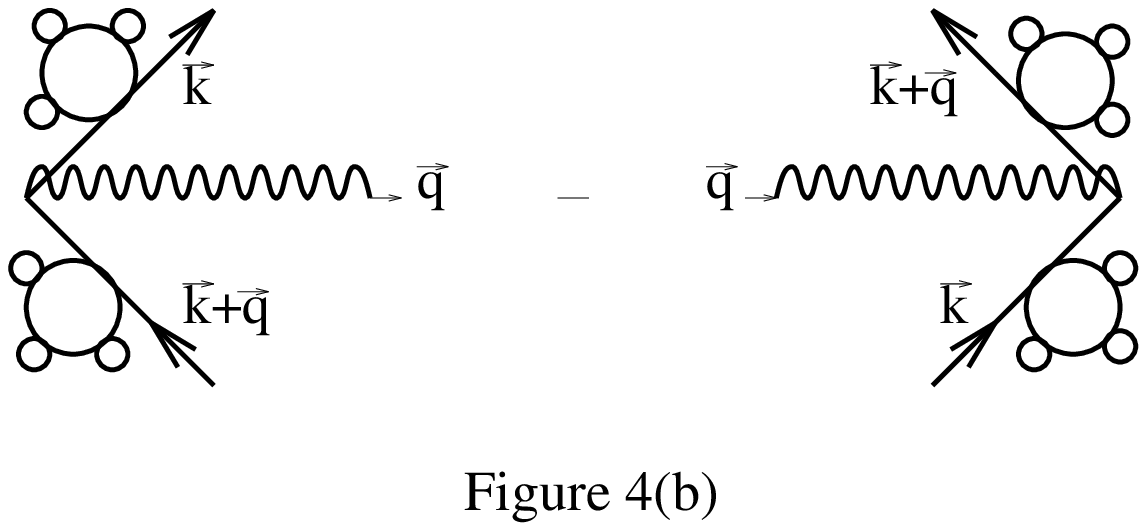}
\epsfig{file=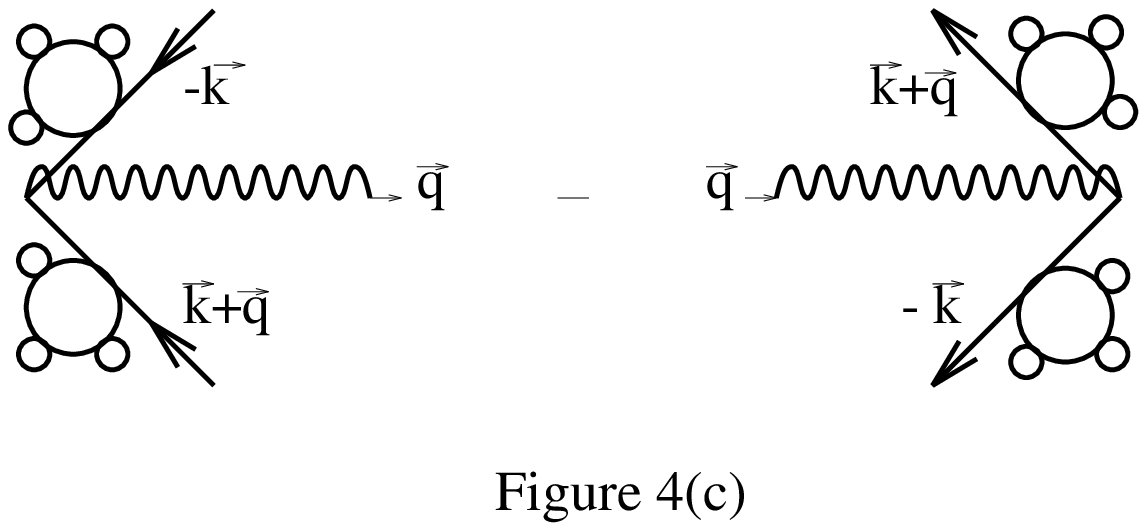}
\caption{Processes that contribute to the photoproduction rate. The pion
propagators are the full large N-resummed propagators.}
\epsfig{file=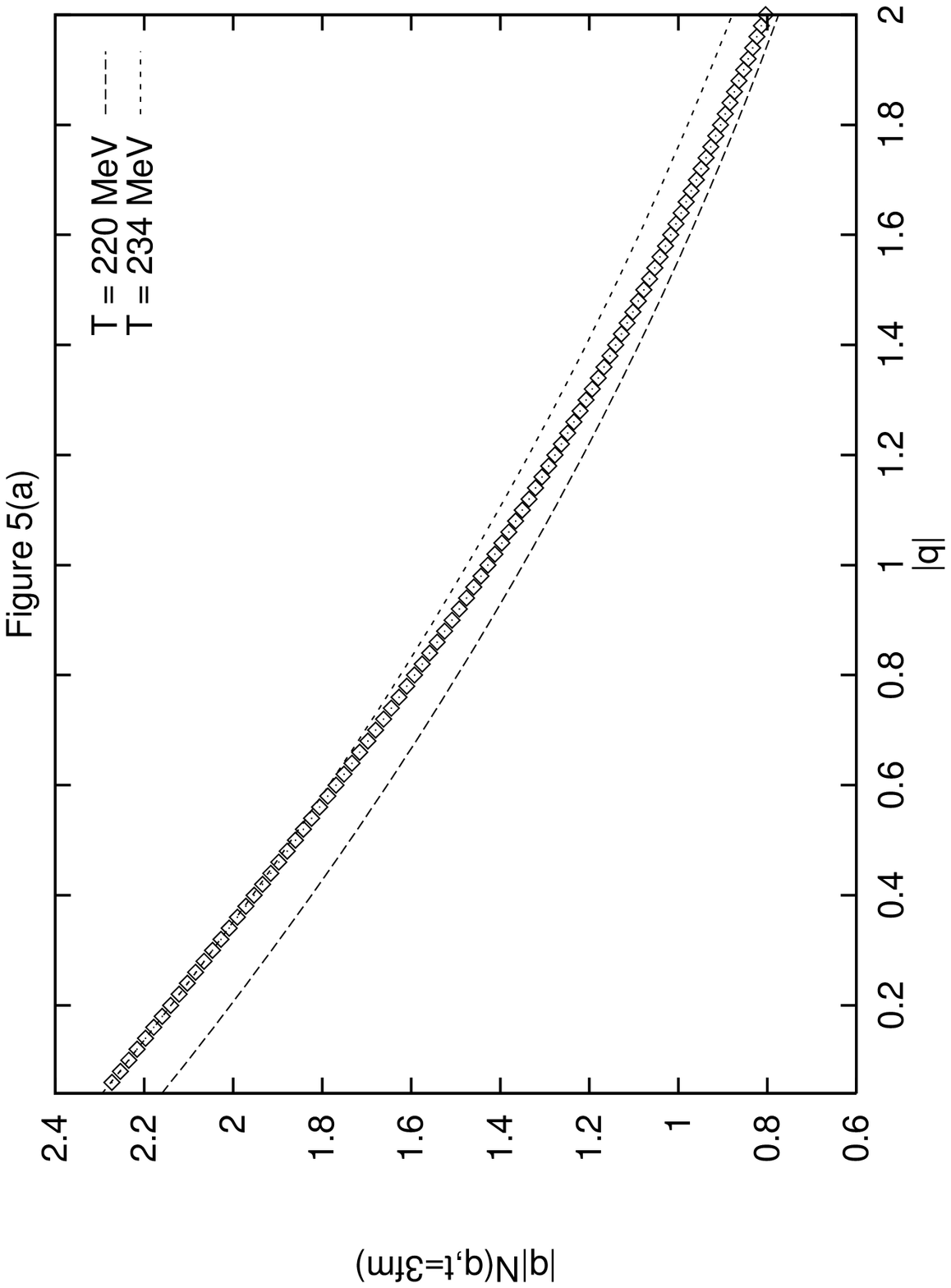}
\caption{(a) Total photon yield per unit volume $(2\pi)^3 |q|\frac{d(\langle
N(t) \rangle/\Omega)}{d^3q}$ in units of $\mbox{fm}^{-1}$ vs. $|q|$ in 
units of 200 Mev, at time $t=$3fm/c, for initial photon occupation in 
equilibrium at temperature $T_i=220 Mev$.}
\addtocounter{figure}{-1}
\epsfig{file=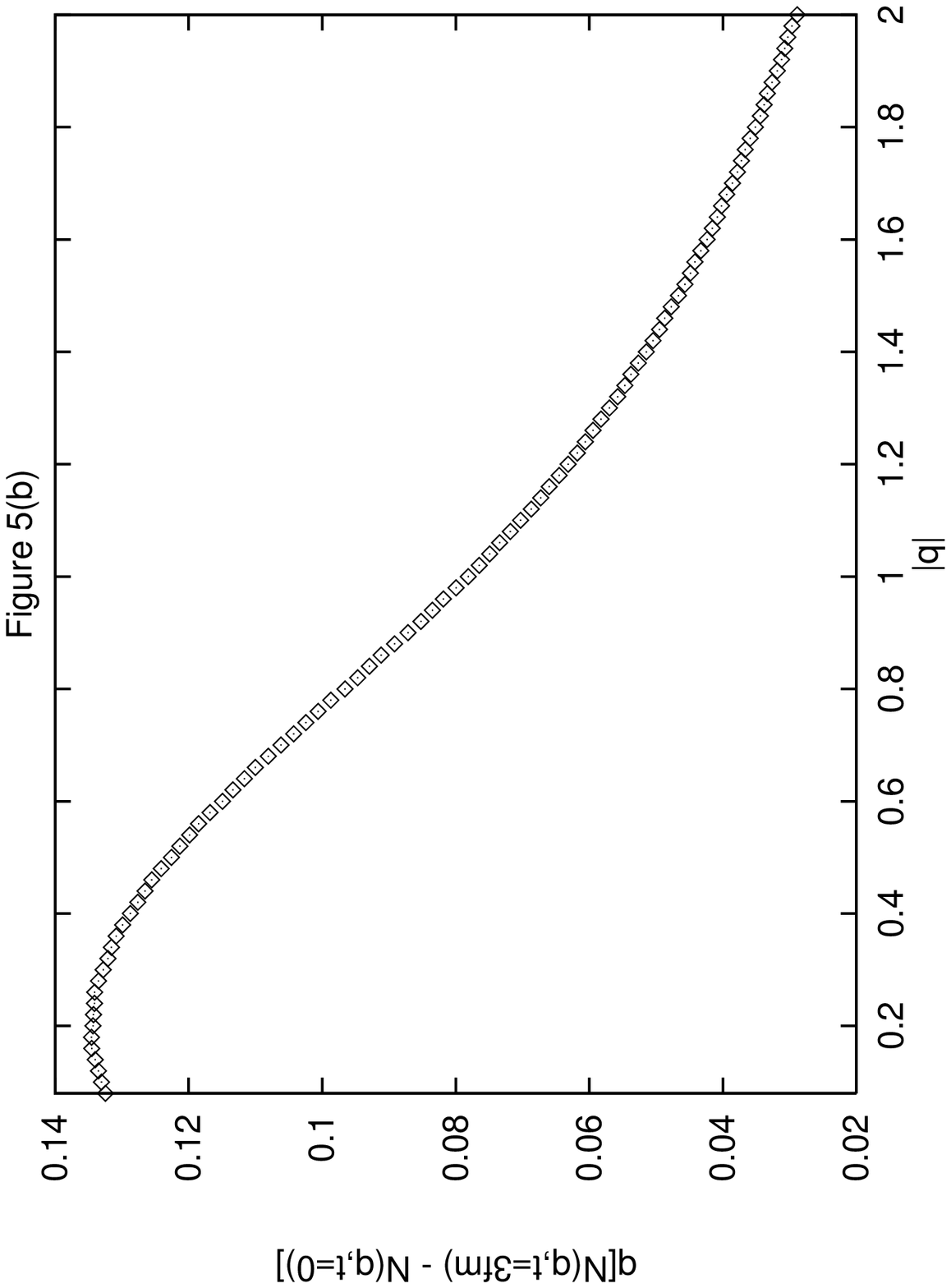}
\caption{(b) Total photon yield per unit volume after subtracting the 
thermal background at $t=0$ same units as in figure (5.a).}
\epsfig{file=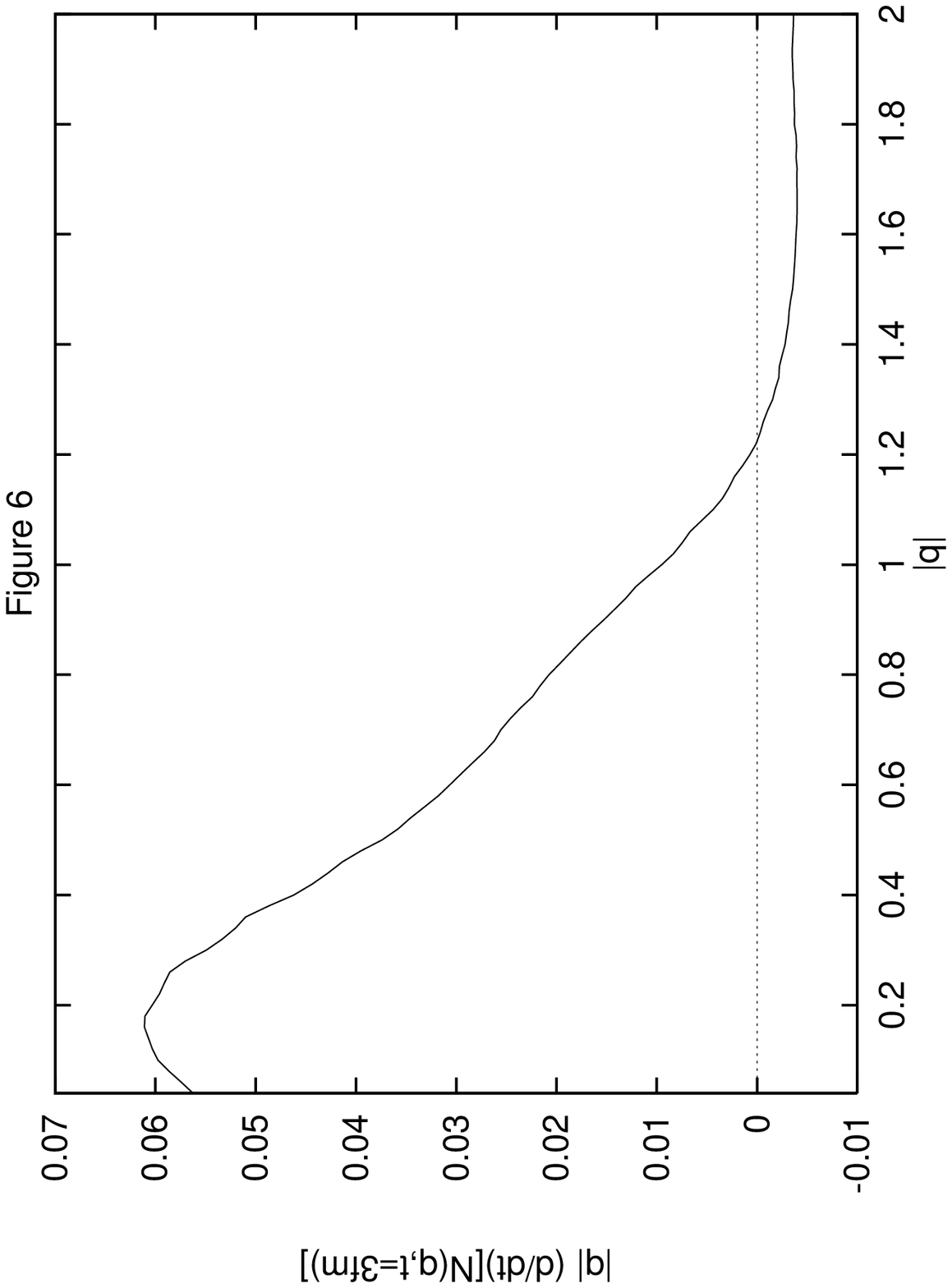}
\caption{The invariant photoproduction rate $(2\pi)^3 |q|\frac{dR(t)}{d^3q}$
in units of $\mbox{fm}^{-2}$ vs.  $|q|$ in units of 200 Mev at time 3 
fm/c, for initial temperature $T_i=220 Mev$.}
\epsfig{file=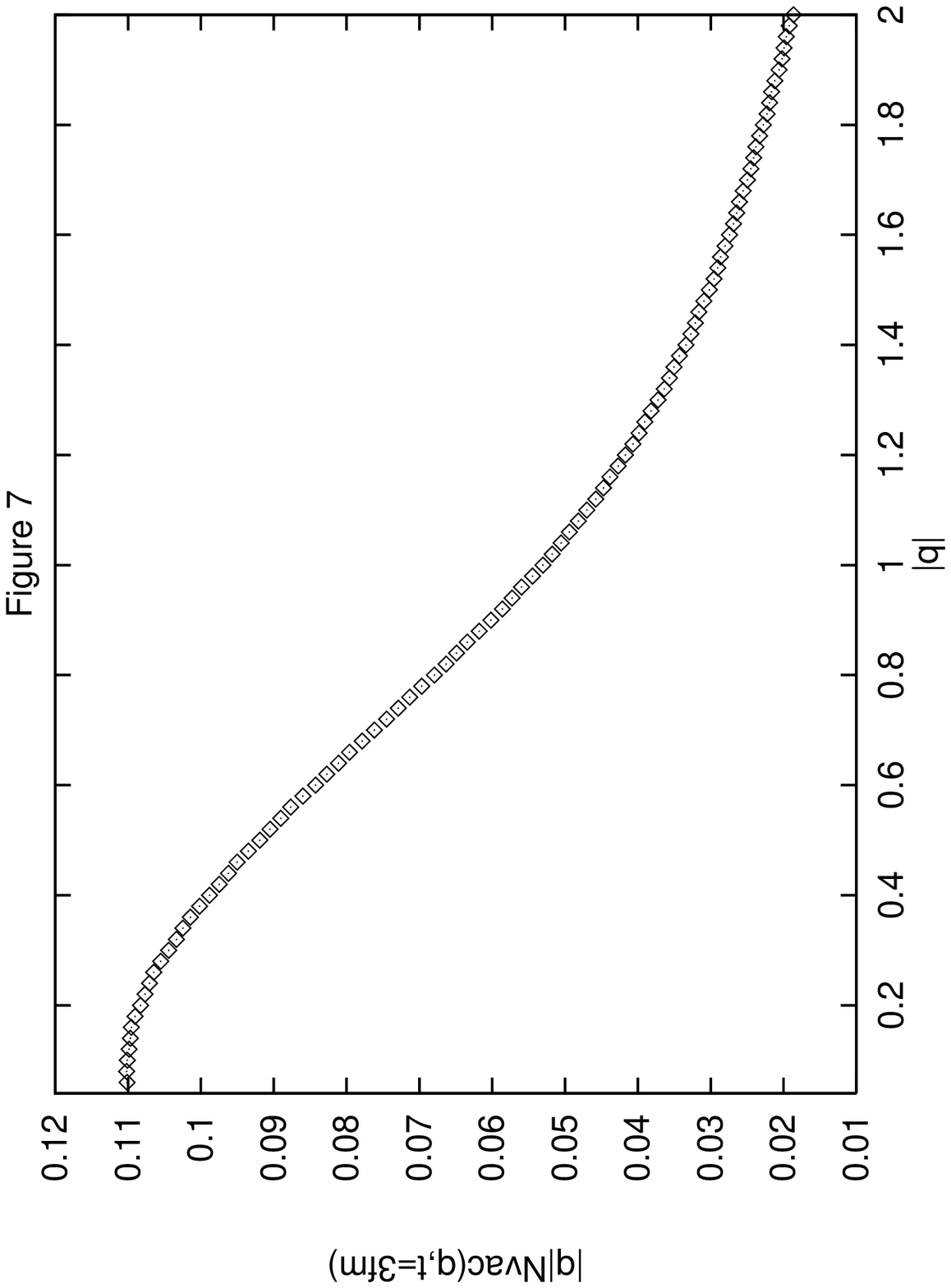}
\caption{Total photon yield per unit volume $(2\pi)^3 |q|\frac{d(\langle N(t)
\rangle/\Omega)}{d^3q}$ in units of $\mbox{fm}^{-1}$ vs. $|q|$ in units 
of 200 Mev, at time $t=$3fm/c, for ``vacuum'' initial photon occupation,
$N_q(0)=0$.}
\epsfig{file=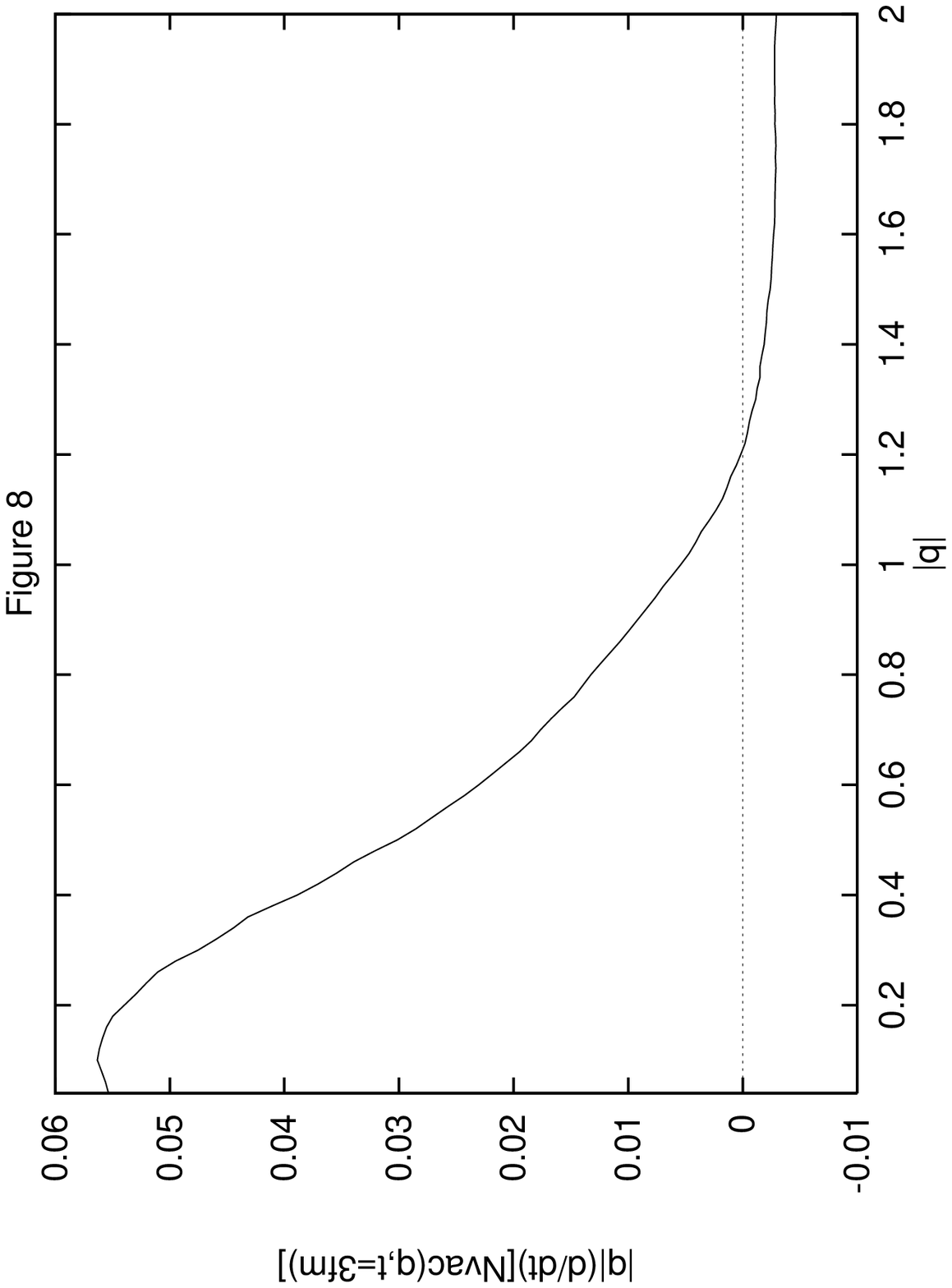}
\caption{The invariant photoproduction rate $(2\pi)^3 |q|\frac{dR(t)}{d^3q}$
in units of $\mbox{fm}^{-2}$ vs.  $|q|$ in units of 200 Mev at time 3 
fm/c, for ``vacuum'' initial photon occupation, $N_q(0)=0$.}
\epsfig{file=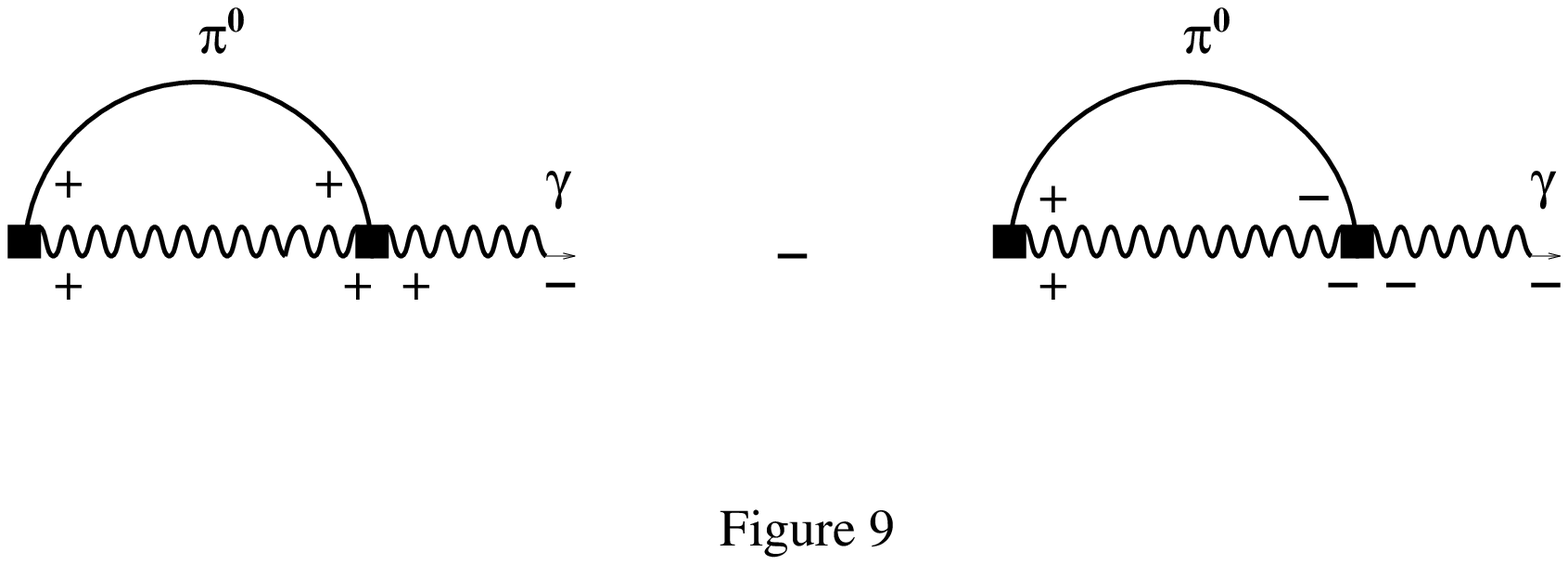}
\caption{Lowest order contribution to photoproduction from the anomalous 
decay of the neutral pion via ${\cal{L}}_A$ in (\ref{anomalousvertex}).}
\end{figure}


\newpage

\begin{thebibliography}{99}
\bibitem{anselm1} A. Anselm, Phys. Letters B217, 169 (1989); A. Anselm and
M. Ryskin,Phys. Letters B226, 482 (1991).
\bibitem{blaizot} J. P. Blaizot and A. Krzywicki, Phys. Rev. D46, 246 (1992);
Acta Phys.Polon.27, 1687-1702 (1996)
and references therein.
\bibitem{bjorken1} J. D. Bjorken, Int. Jour. of Mod. Phys. A7, 4189 (1992);
Acta 
Physica Polonica B23, 561 (1992). See also J. D. Bjorken's contribution
to the proceedings of the ECT workshop on Disoriented Chiral Condensates,
available at http://www.cern.ch/WA98/DCC.
\bibitem{kowalski} K. L. Kowalski and C. C. Taylor, ``Disoriented Chiral
Condensate: A white paper for the Full Acceptance Detector, CWRUTH-92-6.
\bibitem{bjorken2} J. D. Bjorken, K. L. Kowalski and C. C. Taylor: ``Observing
Disoriented Chiral Condensates'', (SLAC-CASE WESTERN preprint 1993)
 hep-ph/9309235 ; ``Baked
Alaska'', (SLAC-PUB-6109) (1993).
\bibitem{revs} For recent reviews on the subject, see: K. Rajagopal, in {\em Quark Gluon Plasma 2},
ed. R. Hwa, World Scientific (1995); S. Gavin, Nucl. Phys. A590 (1995),
163c; J. P. Blaizot and A. Krzywicki,hep-ph/9606263 (1996).  
\bibitem{bjor} J. D. Bjorken, Phys. Rev. D27, 140, (1983).
\bibitem{book1} L. P. Csernai, ``Introduction to Relativistic Heavy Ion
Collisions'', (John Wiley and Sons, England, 1994).
\bibitem{book2} C. Y. Wong, ``Introduction to High-Energy Heavy Ion
Collisions'', (World Scientific, Singapore, 1994).
\bibitem{anselm2} A. Anselm and Myron Bander, ``On the Distribution of
Neutral and Charged Pions through the Production of a Classical Pion Field''
(Irvine preprint UCI-TR-93-32). 
\bibitem{lattes} C. M. G. Lattes, Y. Fujimoto and S. Hasegawa, Phys. Rep. 65,
151 (1980); G. J. Alner et al Phys. Rep. 154, 247 (1987).
\bibitem{wilraj1} K. Rajagopal and F. Wilczek, Nucl. Phys. B379, 395 (1993).
\bibitem{wilraj2} K. Rajagopal and F. Wilczek, Nucl. Phys. B404, 577
 (1993)
\bibitem{boysinlee} D. Boyanovsky, D.-S. Lee and A. Singh, Phys. Rev. D48,
800, (1993).
\bibitem{gavin} S. Gavin, A. Goksch and R. D. Pisarski, Phys. Rev. Lett., 72,
2143 (1994); S. Gavin and B. Muller, Phys. Lett. B329, 486 (1994).
\bibitem{randrup} J. Randrup, Phys. Rev. Lett. 77, (1996), LBL report
38125 (1995); 39328 (1996); hep-ph/9611228 (1996); hep-ph/9612453.
\bibitem{huangwang} Z. Huang  and X. N. Wang, Phys. Rev. D49, 4335 (1994).
\bibitem{cooper} F. Cooper, Y. Kluger, E. Mottola and J. P. Paz,
Phys. Rev. D51, 2377 (1995); Y. Kluger, F. Cooper, E. Mottola, J. P. Paz
and A. Kovner, Nucl. Phys. A590, 581c (1995); 
M. A. Lampert, J. F. Dawson and F. Cooper,
Phys. Rev. D54, 2213-2221 (1996), F. Cooper, Y. Kluger and E. Mottola,
Phys. Rev. C 54, 3298 (1996).
\bibitem{boyan} D. Boyanovsky, H.J. de Vega and R. Holman, Phys. Rev. D51, 734 (1995).
\bibitem{goity} P. Gerber, H. Leutwyler and J.L. Goity,  Phys.Lett.B246,
513,1990.  J.L. Goity and H. Leutwyler Phys.Lett.B228, 517,1989.
\bibitem{minimax} J. D. Bjorken, ``T864 (Minimax): A search for
Disoriented Chiral Condensate at the Fermilab Collider, hep-ph/9610379
(1996). (See also the homepage at http://fnmine.fnal.gov).
\bibitem{wa98} See the homepage of the WA98 collaboration at CERN:
http://www.cern.ch/WA98/DCC.  
\bibitem{alam} J. Alam, S. Raha, B. Sinha, Phys. Reports, 273, 243-362 (1996);
H. Meyer-Ortmanns, Rev.Mod.Phys. 68, 473-598 (1996).
\bibitem{kapvis} J. I. Kapusta and A. P. Vischer 
 ``Dynamical formation of disoriented chiral condensates''
 nucl-th/9605023 (1996).  
\bibitem{wang} Z. Huang and X.- N. Wang, Phys. Lett. B 383, 457 (1996)
\bibitem{fein} E. L. Feinberg, Nuovo Cimento 34A, 39 (1976).
\bibitem{toimela} L. McLerran and T. Toimela, Phys. Rev. D31, 545 (1985). 
\bibitem{rusk} P. V. Ruuskanen, in ``Particle Production in Highly 
Excited Matter'', ed. H. Gutbrod, (Plenum, N.Y.1993); Nucl.Phys. A 544,
169c (1995). 
\bibitem{kapustagale} C. Gale and J. Kapusta, Phys. Rev. C 38, 2659 
(1988).
\bibitem{lichard} J. Kapusta, P. Lichard and D. Seibert, Phys. Rev. D44, 2774 (1991).
\bibitem{zahed} J. V. Steele, H. Yamagishi and I. Zahed, Phys. Lett. B 384, 255 (1996).
\bibitem{schwinger} J. Schwinger, J. Math. Phys. 2, 407 (1961);
L. V. Keldysh, JETP 20, 1018 (1965); K. T. Mahanthappa, Phys. Rev. 126, 329
(1962); P. M. Bakshi and K. T. Mahanthappa, J. Math Phys. 41, 12 (1963).
\bibitem{chou} K. Chou, Z. Su, B. Hao and L. Yu, Phys. Rep. 118, 1 (1985);
A. Niemi and G. Semenoff, Ann. of Phys. (NY) 152, 105 (1984); N. P. Landsmann
and C. G. van Weert, Phys. Rep. 145, 141 (1987); E. Calzetta and B. L. Hu,
Phys. Rev. D 41, 495 (1990); {\em ibid} 37, 2838 (1990); J. P. Paz,
Phys. Rev. D 41, 1054 (1990); {\em ibid} D 42, 529 (1990).
\bibitem{kapusta} J. Kapusta, `Finite Temperature Field Theory', Cambridge
Univ. Press. (1989).
\bibitem{cooper2} F. Cooper and E. Mottola, Mod. Phys. Lett. A 2, 635 (1987);
F. Cooper, S. Habib, Y. Kluger, E. Mottola, J. P. Paz, P. R. Anderson,
Phys. Rev. D50, 2848 (1994); F. Cooper, S.-Y. Pi and P. N. Stancioff,
Phys. Rev. D34, 3831 (1986); F. Cooper, Y. Kluger and E. Mottola
hep-ph/9604284.
\bibitem{strongfields}  F. Cooper, J. M . Eisenberg, Y. Kluger, E. Mottola,
B. Svetitsky, Phys. Rev. Lett. 67, 2427 (1991); F. Cooper in ``Particle
Production in Highly Excited Matter'', Proceedings of the NATO ASI. ed. by
H. Gutgrod and J. Rafelski (Plenum, New York 1993). F. Cooper, J. M. Eisenberg,
Y, Kluger, E. Mottola, B. Svetitsky, Phys. Rev. D48, 190 (1993); See also
J. M. Eisenberg and Y. Kluger contributions in ``Hot and Dense Nuclear Matter''
Ed. by W. Greiner, H. Stocker and A. Gallmann (Nato Asi Series Plenum Press
1994) (pages 333 and 635).
\bibitem{boyan2} D. Boyanovsky, H. J. de Vega and
R. Holman, Phys. Rev. D49, 2769 (1994);
D. Boyanovsky, H. J. de Vega, R. Holman, D. S. Lee and
A. Singh, Phys. Rev. D51, 4419 (1995); D. Boyanovsky, H.J. de Vega, R. Holman,
J.F.J. Salgado hep-ph/9608205, to be published in Phys.Rev.D.
\bibitem{donoghue} J. F. Donoghue, E. Golowich and B. R. Holstein,
``Dynamics of the Standard Model'', Cambridge Univ. Press 1992.
\bibitem{boylawrie} D. Boyanovsky, I.D. Lawrie, D.S. Lee, Phys.Rev. D54,
4013-4028, (1996).
\bibitem{boygauge} D. Boyanovsky, D. Brahm, R. Holman and D.-S. Lee, 
Phys. Rev. D54, 1763 (1996). 
\bibitem{note2} The two terms in (\ref{ndot1}), give
exactly identical contributions to the emission rate. One might worry that the
Heisenberg operators $\vec{J}_T$ and $\vec{\Phi}_T$ in eq.(\ref{ndot2}) might
not commute in the full theory. However with an adiabatic switch-on of the
interactions at some $t_0\rightarrow -\infty$, when the fields are free, then
\[
[\vec{J}_T(t),{\vec{\Phi}}_T(t)]=
U^{-1}(t,t_0)[\vec{J}_T(t_o),{\vec{\Phi}}_T(t_0)]U(t,t_0).
\]
But the commutator on the RHS is zero for free fields, and is preserved at all
times by the unitarity of the theory.
\bibitem{meden}  D. B. Tran Thoai, H. Hang, Phys. Rev. B47, 3574 (1993);
V. Meden, C. Wohler, J. Frick and K. Schonhammer, Phys. Rev. B52, 5624 (1995);
H. Haug and A. P. Jauho, ``Quantum Kinetics in Transport and Optics of
Semiconductors'', (Springer, Berlin 1996).
\bibitem{kinetics} D. Boyanovsky, M. D'Attanasio, H. J. de Vega, R. Holman 
and S. Prem Kumar, in preparation.
\end{thebibliography}
\end{document}